\documentclass[preprint,amsmath,amssymb,aps,nofootinbib,showpacs]{revtex4}
\usepackage{graphicx}
\usepackage{bm}
\usepackage{subfigure}
\usepackage{amsmath}

\newcommand{\epl}{Europhys. Lett.\ }

\newcommand{\pla}{Phys. Lett. A\ }
\newcommand{\jpa}{J. Phys. A\ }
\newcommand{\jpb}{J. Phys. B\ }
\newcommand{\etal}{{\em et al.}}
\newcommand{\e}{\mbox{e}}

\newcommand{\UQ}{ARC Centre of Excellence for Quantum-Atom Optics, 
School of Physical Sciences, University of Queensland, Brisbane, 
QLD 4072, Australia.}
\begin{document}
\title{Quantum dynamics of a four-well Bose-Hubbard model with two different tunneling rates}

\author{C.V. Chianca and M.K. Olsen}
\affiliation{\UQ}
%-----------------------------------------------------------------------
\date{\today}
%------------------------------------------------------------------------

\begin{abstract}

We consider a theoretical model of a four-mode Bose-Hubbard model consisting of two pairs of wells 
coupled via two processes with two different rates. The model is naturally divided into two subsystems with strong intra-system coupling and much weaker coupling between the two subsystems and has previously been introduced as a model for Josephson heat oscillations by Strzys and Anglin [\pra {\bf 81}, 043616 (2010) ].
We examine the quantum dynamics of this model for a range of different initial conditions, in terms of both the number distribution among the wells and the quantum statistics. We find that the time evolution is different to that predicted by a mean-field model and that this system exhibits a wide range of interesting behaviours. We find that the system equilibriates to a maximum entropy state and is thus a useful model for quantum thermalisation.
As our model may be realised to a good approximation in the laboratory, it becomes a candidate for experimental investigation. 

\end{abstract}

  % check PACS

\maketitle

\section{Introduction}

Bose-Einstein condensates (BECs) of weakly interacting dilute gases have long been recognised as a valuable tool for the exploration of the dynamics of non-equilibrium many-body physics. Experimental investigations of BEC have provided a whole new toolbox  for the study of quantum mechanics in mesoscopic systems.  A recent development is the proposal by Strzys and Anglin to use a four-mode Bose-Hubbard model with greatly differing tunneling rates as a model for the investigation of mesoscopic thermodynamics, in particular with regard to the transport of heat~\cite{Anglin}. Their analysis is motivated by the fact that, microscopically, heat is energy stored in degrees of freedom whose evolution is too quick to perceive or control on a  macroscopic time scale. The authors performed an analytical analysis of the system, based on the treatment given to a two well model by Milburn \etal~\cite{Joel}. In fact, their model consists of two of the systems considered by Milburn \etal, with a weaker coupling between these.

In this work we will not focus on the analogy Strzys and Anglin make between slow Josephson oscillations and second sound~\cite{Nozieres}, but instead will investigate the quantum dynamics and statistics of this system. We find that there is a wealth of complex behaviour, very little of which is predicted by mean-field or linearised Bogoliubov type theories. We find that, by going beyond linearised analyses and using the fuller, semi-quantum truncated Wigner approximation~\cite{Wminus1}, we are able to investigate the relaxation of the system to equilibrium. By constructing a measure which behaves very much like von Neumann entropy and which is potentially measurable in the laboratory, we show how the four-well model can make a contribution to the study of thermalisation in isolated quantum systems~\cite{thermal3}.

\section{Physical model and Hamiltonian}
\label{sec:model}

\begin{figure}
\begin{center}
\includegraphics[width=0.45\columnwidth]{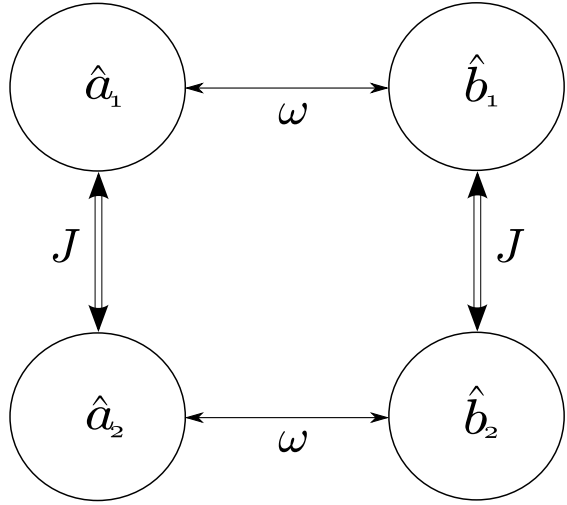}
\end{center}
\caption{Schematic of our four-mode Bose-Hubbard system. The $\hat{a}_{i}$ and $\hat{b}_{i}$ are the bosonic annihilation operators for each mode, while $J$ and $\omega$ represent the coupling rates between the modes.  In this article, we always set $\omega=0.1J$ and $J=1$, which sets the units of time.}
\label{fig:fourmode}
\end{figure}

Extending the standard procedure for two wells~\cite{Joel}, we consider a four well potential with an independent condensate in each of the four wells at the beginning of our investigations. The Hamiltonian for a condensate in an external trapping potential, $V_{ext}(\vec{r})$, may be written as
\begin{equation}
\hat{\cal{H}}=\int{ d \vec{r} \left[ \frac{\hbar^2}{2m} \nabla \hat{\psi}^{\dagger} \cdot \nabla \hat{\psi} +V_{ext}(\vec{r}) + \hbar U_0\hat{\psi}^{\dagger} \hat{\psi}^{\dagger} \hat{\psi} \hat{\psi}   \right] } , 
\label{eq:hamiltonian}
\end{equation}
where $\hat{\psi}$ is the field operator for the condensate, and the non-linear interaction parameter is $U_0=2\pi a \hbar / m$,  where $a$ is the s-wave scattering length describing two-body collisions within the condensate, and $m$ is the atomic mass. In the case where the external potential provides a four well confinement for the condensate, we may simplify the above Hamiltonian by making use of the four-mode approximation. At zero temperature all atoms in the system are condensed and if the ground state energies of the condensate in the four single (and separate) wells are sufficiently separated from the energies of the condensate in all other excited single particle states, transitions to or from the two modes of interest and these higher lying states can be neglected. We may then expand the field operator as
\begin{equation}
\hat{\psi}(\vec{r})\approx  \displaystyle\sum\limits_{i=1}^{2}\left(\phi^{L}_{i}(\vec{r}) \hat{a}_{i} + \phi^{R}_{i}(\vec{r})\hat{b}_{i}\right) ,
\label{eq:4modeoperators}
\end{equation}
where $\hat{a}_{i}$ and $\hat{b}_{i}\: (i=1,2)$ are bosonic annihilation operators in each of the wells, and the $\phi^{L/R}_{i}$ are the ground state spatial wave functions of the condensate in wells on the left and right side, as shown schematically in Fig.~\ref{fig:fourmode}.

Using this in Eq.~(\ref{eq:hamiltonian}), we find an effective Hamiltonian
\begin{eqnarray}
\hat{\cal{H}}_{eff} &=& \displaystyle\sum\limits_{i=1}^{2} \left( \hbar E_{i}^{L} \hat{a}_{i}^\dagger \hat{a}_{i}+ \hbar E_{i}^{R} \hat{b}_{i}^\dagger \hat{b}_{i}\right.\nonumber\\
& & \left. + \hbar\chi \hat{a}_{i}^\dagger \hat{a}_{i}^\dagger \hat{a}_{i} \hat{a}_{i}+\hbar\chi \hat{b}_{i}^\dagger \hat{b}_{i}^\dagger \hat{b}_{i} \hat{b}_{i}\right)\nonumber\\
& &
-\hbar J \left(\hat{a}_{1}^\dagger \hat{a}_{2}+\hat{a}_{2}^\dagger \hat{a}_{1}+\hat{b}_{1}^\dagger \hat{b}_{2}+\hat{b}_{2}^\dagger \hat{b}_{1} \right)\nonumber\\
& & -\hbar\omega\left(\hat{a}_{1}^\dagger \hat{b}_{1}+\hat{b}_{1}^\dagger \hat{a}_{1}+\hat{a}_{2}^\dagger \hat{b}_{2}+\hat{b}_{2}^\dagger \hat{a}_{2}\right),
\label{eq:effhamiltonian}
\end{eqnarray}
where we have neglected the spatial overlap of the different well densities. The single well bound state energies, $E_{i}^{L/R}$, are
\begin{equation}
E_{i}^{L/R}=\frac{1}{\hbar}\int{d\vec{r}~ (\phi_{i}^{L/R})^{\ast}(\vec{r}) \left(\frac{-\hbar^2}{2m}\nabla^2 + V_{ext}(\vec{r}) \right) \phi_{i}^{L/R}(\vec{r}) }.
\label{eq:wellenergies}
\end{equation}
$J$, the tunnel coupling on each side of the system, is
\begin{equation}
J = \frac{-1}{\hbar}\int{d\vec{r}~ (\phi_{1}^{L/R})^{\ast}(\vec{r}) \left(\frac{-\hbar^2}{2m}\nabla^2 + V_{ext}(\vec{r}) \right) \phi_{2}^{L/R}(\vec{r}) } ,
\label{eq:Jdeff}
\end{equation}
while $\omega$, the tunnel coupling between the left and right subsystems, is
\begin{equation}
\omega = \frac{-1}{\hbar}\int{d\vec{r}~ (\phi_{i}^{L})^{\ast}(\vec{r}) \left(\frac{-\hbar^2}{2m}\nabla^2 + V_{ext}(\vec{r}) \right) \phi_{i}^{R}(\vec{r}) } .
\label{eq:wdeff}
\end{equation}
The effective non-linear interaction term is
\begin{equation}
\chi = U_0 \int{d\vec{r}~ \vert \phi_{i}^{L/R}(\vec{r}) \vert^4 } .
\label{eq:interactioneqn}
\end{equation}
We set the single well bound state energies equal because we will consider only a symmetric potential where we can set $E_L=E_R=0$. We note here that, while  Strzys and Anglin began their dynamical investigations from the ground state and provided a periodic tilt to the potentials to excite the dynamics, we leave the potential unperturbed and excite the dynamics via differences in the initial populations of the wells. 

We parametrise time by setting $J=1$, so that dimensionless time as displayed in the results will be in units of $Jt$. We will investigate the effects of changing $\chi$ and the initial distribution of atoms in the wells. Because we use a quantum phase space method, we may also change the quantum statistics of the initial states in each well. In this work we will investigate the dynamics arising from initial Fock and coherent states~\cite{Danbook}. We will always use a total average atom number of $N_{T}=10^{4}$.

\section{Theoretical methods}
\label{sec:methods} 

For the numerical investigation of many-body interacting quantum systems which are too large for master equation methods, the preferred first option is the positive-P representation~\cite{P+}, which allows for an exact mapping from the type of Hamiltonian used here to stochastic differential equations. However, in cases where the system is undamped and has a high $\chi^{(3)}$ nonlinearity, it tends to become unstable after very short times~\cite{Steel}. 
As this is the case here, we will perform our investigations using stochastic integration in the truncated Wigner
representation~\cite{Wminus1,Steel}, which enables us to capture the
majority of the quantum features of the system as long as the Wigner
pseudoprobability distribution is positive. There is no reason to believe this is not the case in any of the investigations we perform here except in the representation of initial Fock states, where we will use an approximation which is justified below. The truncated Wigner representation also has the huge operational advantage of remaining stable over relatively long integration times. The truncated Wigner representation has been shown to be accurate for the investigation of a range of condensate dynamics~\cite{RCsuper,Wigsuper,TW3}, and we have also found that its predictions are accurate for twin-well dynamics, so expect it to be accurate here.

To find the appropriate equations, we begin by using the operator
correspondences~\cite{QNoise}
\begin{eqnarray}
\hat{a}\rho &\leftrightarrow& \left( \alpha +
\frac{1}{2}\frac{\partial}{\partial \alpha^{\ast}}\right) W(\alpha)
\\ \rho \hat{a} &\leftrightarrow& \left( \alpha -
\frac{1}{2}\frac{\partial}{\partial \alpha^{\ast}}\right) W(\alpha)
\\ \hat{a}^{\dagger}\rho &\leftrightarrow& \left( \alpha^{\ast} -
\frac{1}{2}\frac{\partial}{\partial \alpha}\right) W(\alpha) \\ \rho
\hat{a}^{\dagger} &\leftrightarrow& \left( \alpha^{\ast} +
\frac{1}{2}\frac{\partial}{\partial \alpha}\right) W(\alpha),
\label{eq:correspondences}
\end{eqnarray}
to give a generalised Fokker-Planck equation with third-order
derivatives. Although it is possible to map this approximately onto
stochastic differential equations~\cite{nossoEPL}, the numerical
integration of these is extremely unstable, so we will instead use
what is known as the truncated Wigner representation by dropping derivatives of higher than second
order in the Fokker-Planck equation. This leaves an
equation with no diffusion terms for the Wigner pseudoprobability function,
\begin{eqnarray}
\frac {dW}{dt} & = &\left\{ - \left[ \frac {\partial}{\partial
    \alpha_{1}}\left( -2i \chi |\alpha_{1}|^{2} \alpha_{1} + iJ
  \alpha_{2} + i\omega \beta_{1} \right) + \frac {\partial}{\partial
    \alpha_{1}^{\ast}} \left( 2i \chi
  |\alpha_{1}|^{2}\alpha_{1}^{\ast} - iJ \alpha_{2}^{\ast} - i\omega
  \beta_{1}^{\ast} \right) \right. \right. \nonumber \\ 
  &&
  \left. \left. +\frac {\partial}{\partial \alpha_{2}} \left( -2i \chi
  |\alpha_{2}|^{2} \alpha_{2} + iJ \alpha_{1} + i\omega \beta_{2}
  \right) +\frac {\partial}{\partial \alpha_{2}^{\ast}} \left( 2i
  \chi |\alpha_{2}|^{2} \alpha_{2} ^{\ast}- iJ \alpha_{1}^{\ast} -
  i\omega \beta_{2}^{\ast} \right) \right. \right. \nonumber \\ 
  &&
  \left. \left. +\frac {\partial}{\partial \beta_{1}} \left( -2i \chi
  |\beta_{1}|^{2} \beta_{1} + iJ \beta_{2} + i\omega \alpha_{1}
  \right) +\frac {\partial}{\partial \beta_{1}^{\ast}} \left( 2i \chi
  |\beta_{1}|^{2} \beta_{1}^{\ast} - iJ \beta_{2}^{\ast} - i\omega
  \alpha_{1}^{\ast} \right) \right. \right. \nonumber \\ 
  &&
  \left. \left.+\frac {\partial}{\partial \beta_{2}} \left( -2i \chi
  |\beta_{2}|^{2} \beta_{2} + iJ \beta_{1} + i\omega \alpha_{2}
  \right) +\frac {\partial}{\partial \beta_{2}^{\ast}} \left( 2i \chi
  |\beta_{2}|^{2} \beta_{2}^{\ast} - iJ \beta_{1}^{\ast} - i\omega
  \alpha_{2}^{\ast} \right) \right] \right\} W.\nonumber\\
\label{eq:truncFPE}
\end{eqnarray}

The equations of motion for the Wigner variables of this system are then found as
\begin{eqnarray}
\frac {d \alpha_{1}}{dt} &=&
-2i\chi|\alpha_{1}|^{2}\alpha_{1} +iJ\alpha_{2} + i\omega \beta_{1}\nonumber\\
\frac {d \alpha_{2}}{dt} &=&
-2i\chi|\alpha_{2}|^{2}\alpha_{2} +iJ\alpha_{1} + i\omega \beta_{2}\nonumber\\
\frac {d \beta_{1}}{dt} &=&
-2i\chi|\beta_{1}|^{2}\beta_{1} +iJ\beta_{2} + i\omega \alpha_{1}\nonumber\\
\frac {d \beta_{2}}{dt} &=&
-2i\chi|\beta_{2}|^{2}\beta_{2} +iJ\beta_{1} + i\omega \alpha_{2},
\label{eq:Wminus}
\end{eqnarray}
where the $\alpha_{j}$ and $\beta_{j}$ are the Wigner variables
corresponding to $\hat{a}_{j}$ and $\hat{b}_{j}$,
respectively. Classical averages of the Wigner variables correspond to
symmetrically ordered operator expectation values, so that the
necessary reordering must be undertaken before we arrive at solutions
for physical quantities, for which normal ordering is more appropriate. Although Eq.~(\ref{eq:Wminus}) might look
classical, the Wigner variables themselves are drawn from appropriate
distributions for the desired initial states, so that the
stochasticity comes from the initial conditions. The truncated Wigner equations above are solved numerically by taking averages over a large number of stochastic trajectories, with initial conditions drawn from the distributions given below.

The dynamical evolution of this system can depend on the initial quantum state as well as the initial number distribution. In this case, we will investigate two different initial number distributions and two different initial quantum states. Most of our analyses will be performed with half the atoms in each of two diagonally opposite wells and the other two initially vacant, with equal populations in each well also being used to calculate the Josephson frequencies. 
Using the Wigner representation we may easily simulate different initial quantum states~\cite{AxMuzza}. 
To represent the Wigner distribution for a coherent state, $|\alpha\rangle$, where $\hat{a}|\alpha\rangle = \alpha|\alpha\rangle$, the initial conditions are chosen from the distribution
\begin{equation}
\alpha_{W}=\alpha+\frac{1}{2}(\nu_{1}+i\nu_{2}),
\label{eq:Wcoherent}
\end{equation}
where the $\nu_{j}$ are independent Gaussian normal random variables. We easily
see that, as required by the symmetric ordering,
$\overline{|\alpha_{W}|^{2}}=N_{a}+1/2$. For simulations using initial coherent states, we used the open source software package xmds~\cite{xmds}.
Fock states of fixed atom number may be simulated using a Gaussian approximation developed by
Gardiner \etal~\cite{Fudge} and previously used to analyse trapped BEC photoassociation~\cite{Fock}.   A Fock state of fixed atom number, $|N\rangle$,
can be sampled, to a good approximation as long as $N$ is not too
small, by
\begin{equation}
\alpha_{W} = (p+q\nu)\e^{2i\pi\xi},
\label{eq:Focksample}
\end{equation}
where $\xi$ is a random number from the uniform distribution $[0,1)$ and
$q=1/4p$, with
\begin{equation}
p=\frac{1}{2}\left(2N+1+2\sqrt{N^{2}+N}\right)^{1/2}.
\end{equation}
This approximation has been shown to reproduce well the first two
moments for reasonable sizes of $N$~\cite{Fock}, which is all that is
required of a Gaussian distribution. Simulations using initial Fock states were performed in Matlab which, although not as fast as xmds, does have a uniform random number generator. We note here that newer versions of xmds also have a uniform random number generator, but this was not available at the time we ran our simulations.  The initially unoccupied wells have a distribution chosen from Eq.~(\ref{eq:Wcoherent}) with $\alpha=0$, which reproduces the vacuum state.

We note here that we have used these two quantum states not because they are actually what we would necessarily expect the physical quantum state of a condensate to be, but because they are in common usage and serve to show any dynamical differences which may arise from differences in the initial quantum statistics. In fact, other states have been used previously, with Olsen and Plimak~\cite{RCsuper,Wigsuper}
using both squeezed states and states which are sheared in the phase-space~\cite{DCW} to investigate BEC photoassociation.

\section{Quantum and classical dynamics}
\label{sec:quantclass}

\begin{figure}
\begin{center}
\includegraphics[width=0.6\columnwidth]{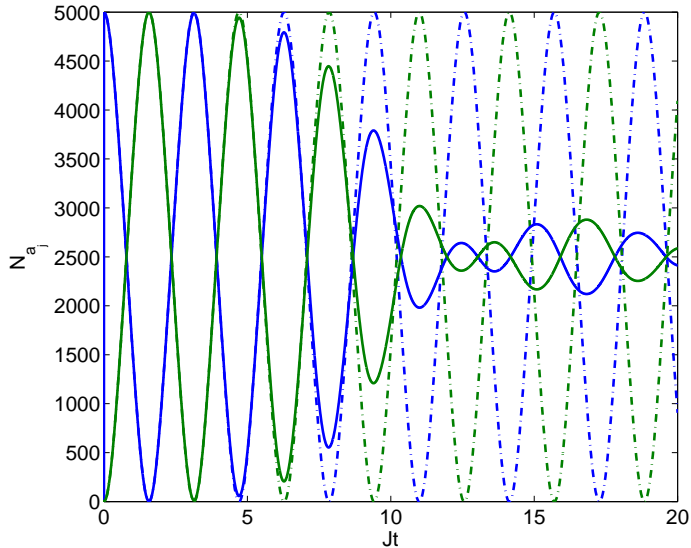}
\end{center}
\caption{(colour online) Populations $Na_{1}$ and $Na_{2}$ with the atoms initially distributed evenly in wells $a_{1}$ and $b_{2}$. The parameters used are $J=1$, $\omega=0.1$ and $\chi=2.5J/N_{T}$, where $N_{T}$ is the total number of atoms, with a value of $10^{4}$. The dash-dotted lines are the classical predictions, while the solid lines are for initial coherent states in the populated wells. The coherent state prediction is the result of $6\times 10^{6}$ stochastic trajectories. The time axes in this and subsequent time domain figures are in dimensionless units.}
\label{fig:Anglinsys}
\end{figure}

When they introduced this system, Strzys and Anglin stated that the full range of quantum dynamics was rich beyond the scope of their paper~\cite{Anglin}. In this work we reveal a small part of this rich dynamics, beginning by exposing the differences from the classical dynamics as predicted in the coupled Gross-Pitaevskii equation (GPE) approach. In their paper, Strzys and Anglin claim that the number of atoms they use ($10^{4}$) is large enough so that the dynamics will stay close to the classical predictions. However, when we use the same parameters as in their paper, with $J=1$, $\omega=0.1$ and $\chi=2.5J/N_{T}$, where $N_{T}$ is the total number of atoms, we see that this is not the case. In Fig.~\ref{fig:Anglinsys}, we compare the classical Gross-Pitaevskii predictions for the populations in wells $a_{1}$ and $a_{2}$ to those found for initial coherent states in the truncated Wigner representation, we see that they are markedly different after the first few oscillations. We note here that, although we  have not used the same initial number distribution as Strzys and Anglin, our result shows that having a relatively large total number of atoms is not sufficient in itself to mean that the dynamics are a small perturbation around the classical predictions.  We also note that, due to the remaining symmetry of the system for these parameters and initial conditions, we only need to show the populations of one of the two wells on one side to demonstrate the full population dynamics.

\begin{figure}
\begin{center}
\includegraphics[width=0.6\columnwidth]{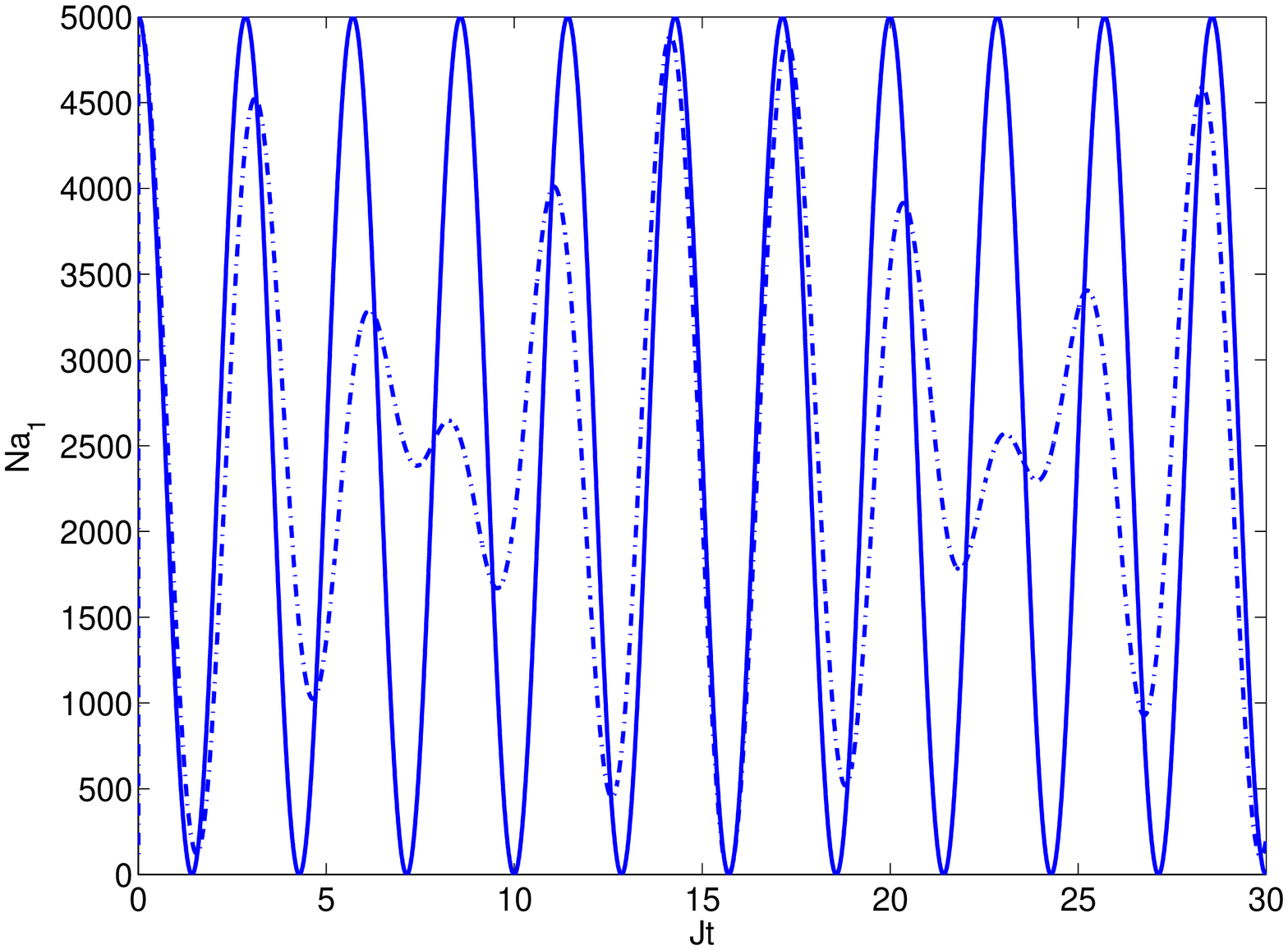}
\end{center}
\caption{(colour online) Population $Na_{1}$ with the atoms initially distributed evenly in wells $a_{1}$ and $b_{2}$. The parameters used are $J=1$, $\omega=0.1$ and $\chi=0.1J/N_{T}$, where $N_{T}$ is the total number of atoms, with a value of $10^{4}$. The solid line is the classical prediction, with the prediction for initial coherent states being identical over this length of time, while the dash-dotted line is for initial Fock states in the populated wells. The coherent state prediction is the result of $2\times 10^{5}$ stochastic trajectories while the Fock state result is the average of $4\times 10^{6}$ trajectories.}
\label{fig:Na1compare}
\end{figure}

In this work, we will use much lower collisional nonlinearities ($\chi$) than that used by Strzys and Anglin, but the dynamics we will investigate will still not be a small perturbation about the classical predictions. 
To illustrate this, as shown in Fig.~\ref{fig:Na1compare}, we take an initial condition with half the atoms in each of wells $a_{1}$ and $b_{2}$ and a nonlinearity of $\chi=0.1J/N_{T}$, with the two different quantum states, and compare these to the prediction of the classical equations.
Over the time shown, the prediction for initial coherent states is almost indistinguishable from that of the GPE approach, although a collapse in the oscillations is seen at longer times. This collapse is normally explained for $\chi^{(3)}$ systems as being due to the fact that different number components of the coherent state superposition oscillate at different rates, so that they eventually become out of phase. Although supersymmetry arguments say that atoms cannot actually be in coherent states~\cite{Cirac}, we would expect the same dynamics for a mixture of number states. The Fock state results, on the other hand, are very different, unlike the case of two coupled wells where they only become markedly different for higher collisional strengths. 

\begin{figure}
\begin{center}
\subfigure{\includegraphics[width=0.45\columnwidth]{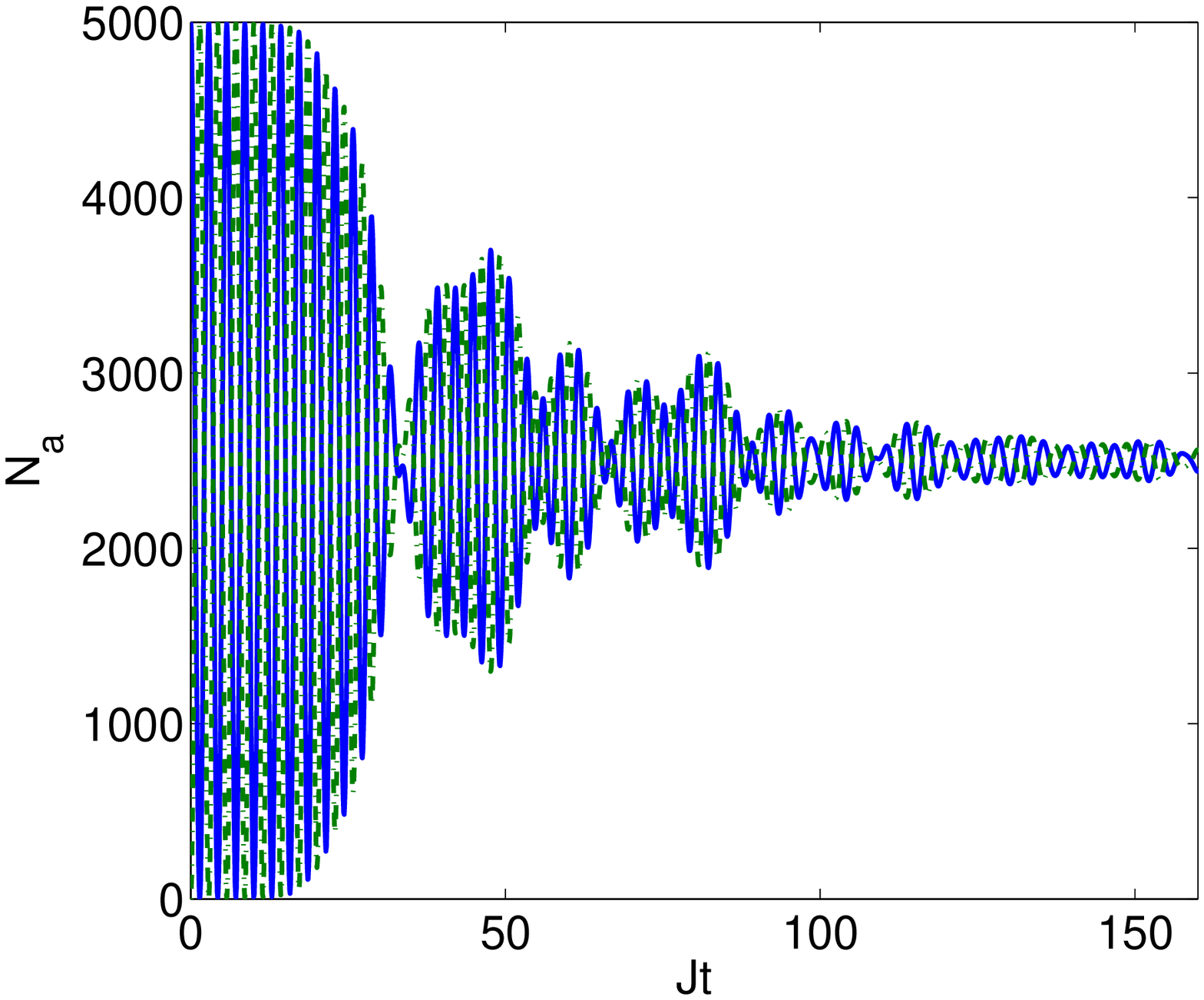}
\label{fig:Coh_05}}
\hspace{8pt}
\subfigure{\includegraphics[width=0.45\columnwidth]{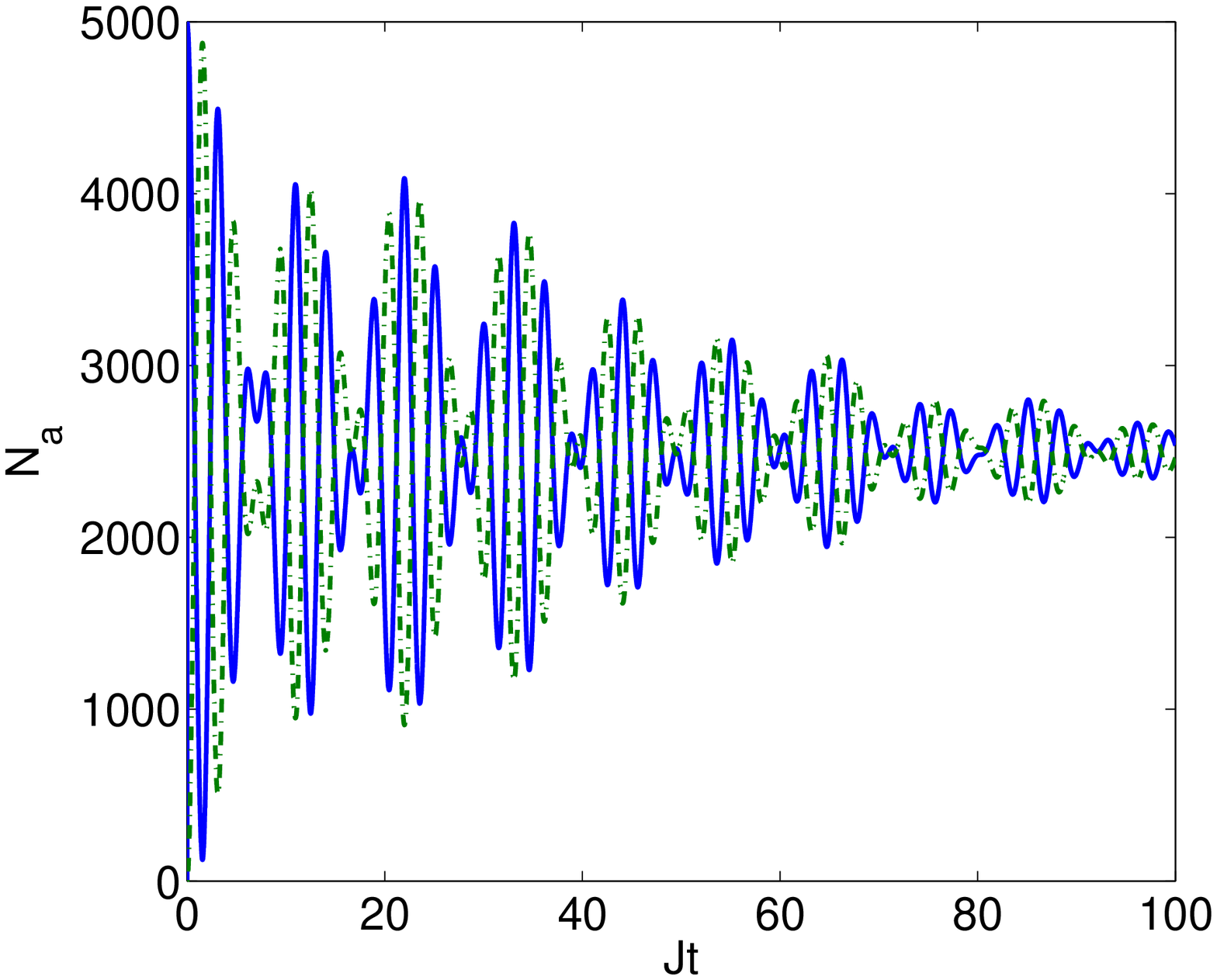}
\label{fig:Fock_05}}
\end{center}
\caption{(colour online) The longer time  population dynamics of wells $a_{1}$ and $a_{2}$ for the same parameters and initial conditions as in Fig.~\ref{fig:Na1compare}, except that the collisional interaction strength is now $\chi=0.5J/N_{T}$. \subref{fig:Coh_05} shows the dynamics for initial coherent states, while \subref{fig:Fock_05} is for initial Fock states. We see that the average number in each well tends to a state where they oscillate about an equal distribution, which did not happen for the lower value of $\chi$.}
\label{fig:collapse05}
\end{figure}

When we increase the collisional interaction to $\chi=0.5J/N_{T}$, we see that the mean-field dynamics change qualitatively for both initial quantum states, as shown in Fig.~\ref{fig:collapse05}. The collapses and partial revivals in the oscillations occur on a time scale inversely proportional to the interaction strength and both initial distributions equilibriate to equal numbers in each of the four wells.  For these initial number distributions, and for values of $\chi \leq 2J/N_{T}$, the classical prediction is for totally regular oscillations, not displaying any of the macroscopic self-trapping predicted in the two-well system~\cite{Joel}.  An analysis of the number and quadrature statistics shows that the atoms in each well evolve away from being in either coherent or Fock states, with more noise in both quadratures and intensity than coherent states, but less noise in the quadratures than for Fock states. Another difference is that we did not see complete revivals in the oscillations for the four-well system. We will give an explanation for their absence in terms of relaxation and the possibility of chaotic behaviour in section~\ref{sec:relax}.
We stress here that this behaviour would not have been observable in the usual linearised analyses, which depend on the accuracy of the GPE type approach as a starting point. The difference between the classical GPE dynamics and those for the two initial quantum states we consider here shows that, for this system, a total number of $10^{4}$ atoms is not sufficient for it to be treated classically, but that both the initial quantum state and the quantum dynamics must be considered to obtain accurate predictions, even for the mean fields.  The time over which we have integrated the equations is different for each initial state for three reasons. The first of these is that xmds is faster than Matlab, the second is that we need more samples to faithfully reproduce Fock states, and the third is that the results with initial coherent states generally take longer to demonstrate qualitative differences in their dynamics. 

The use of phase-space methods also allows us to calculate the dynamics of the Schwinger pseudospin operators~\cite{SU2} adapted to this four-mode system. Given that there is no diagonal tunnelling, that we are examining two linked subsystems, and the symmetry of the initial conditions we consider, it is sufficient to define
\begin{eqnarray}
S_{z}^{a} &=& \hat{a}_{1}\hat{a}_{2}^{\dag}+\hat{a}_{1}^{\dag}\hat{a}_{2},\nonumber\\
S_{y}^{a} &=& -i\left(\hat{a}_{1}\hat{a}_{2}^{\dag}-\hat{a}_{1}^{\dag}\hat{a}_{2}\right),\nonumber\\
S_{z}^{(1)} &=& \hat{a}_{1}\hat{b}_{1}^{\dag}+ \hat{a}_{1}^{\dag}\hat{b}_{1},\nonumber\\
S_{y}^{(1)} &=& -i\left(\hat{a}_{1}\hat{b}_{1}^{\dag}-\hat{a}_{1}^{\dag}\hat{b}_{1}\right),
\label{eq:SU2things}
\end{eqnarray}
with the obvious changes being made to arrive at $S_{z}^{(2)}$ and $S_{y}^{(2)}$. As the $S_{x}$ represent number differences, we have not used them here. 

The $S_{z}$ represent the particle occupation number difference between the single party energy eigenstates of two adjacent wells, and is clearly related to the coherence between these two wells, while the $S_{y}$ represent the momenta between adjacent wells. We also note that we use a different normalisation convention to that used by some authors, who insert a factor of $1/\sqrt{2}$ in front of the expressions. We will also normalise $S_{z}^{(1)}$ by dividing by $\langle \hat{a}_{1}^{\dag}\hat{a}_{1}+\hat{b}_{1}^{\dag}\hat{b}_{1}\rangle$ and $S_{z}^{a}$ by $\langle \hat{a}_{1}^{\dag}\hat{a}_{1}+\hat{a}_{2}^{\dag}\hat{a}_{2}\rangle$, so that the range of possible values lies between $-1$ and $1$, with $0$ representing no occupation number difference of the eigenstates and $\pm 1$ representing the maximum possible. As can be seen in the Hamiltonian, the $S_{z}$ are also proportional to the energy of tunnelling between adjacent wells.
We will now show that the dynamical expectation values of these quantities also depend strongly on the initial quantum states.

\begin{figure}
\begin{center}
\subfigure{\includegraphics[width=0.45\columnwidth]{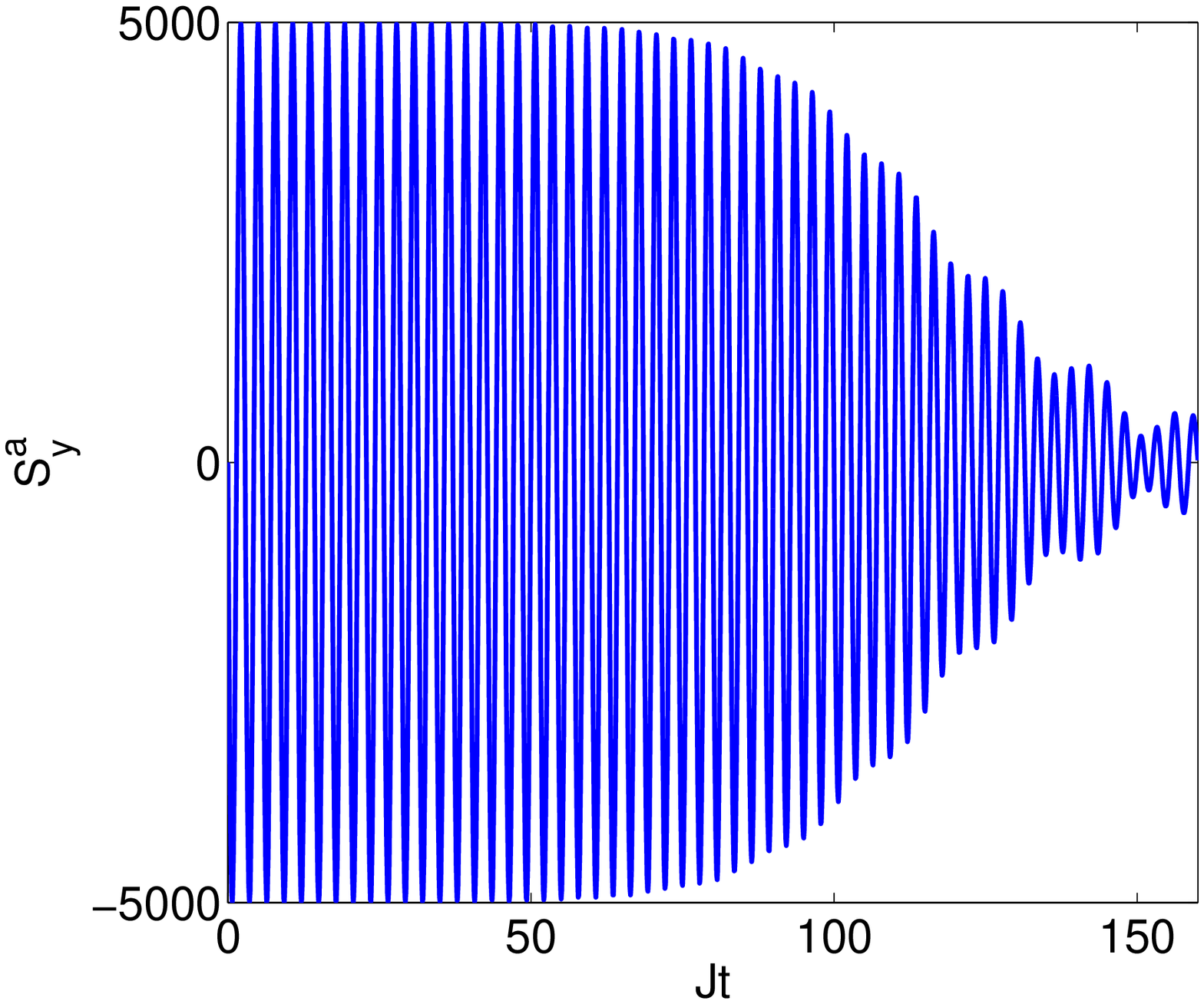}
\label{fig:Sya1c}}
\hspace{8pt}
\subfigure{\includegraphics[width=0.45\columnwidth]{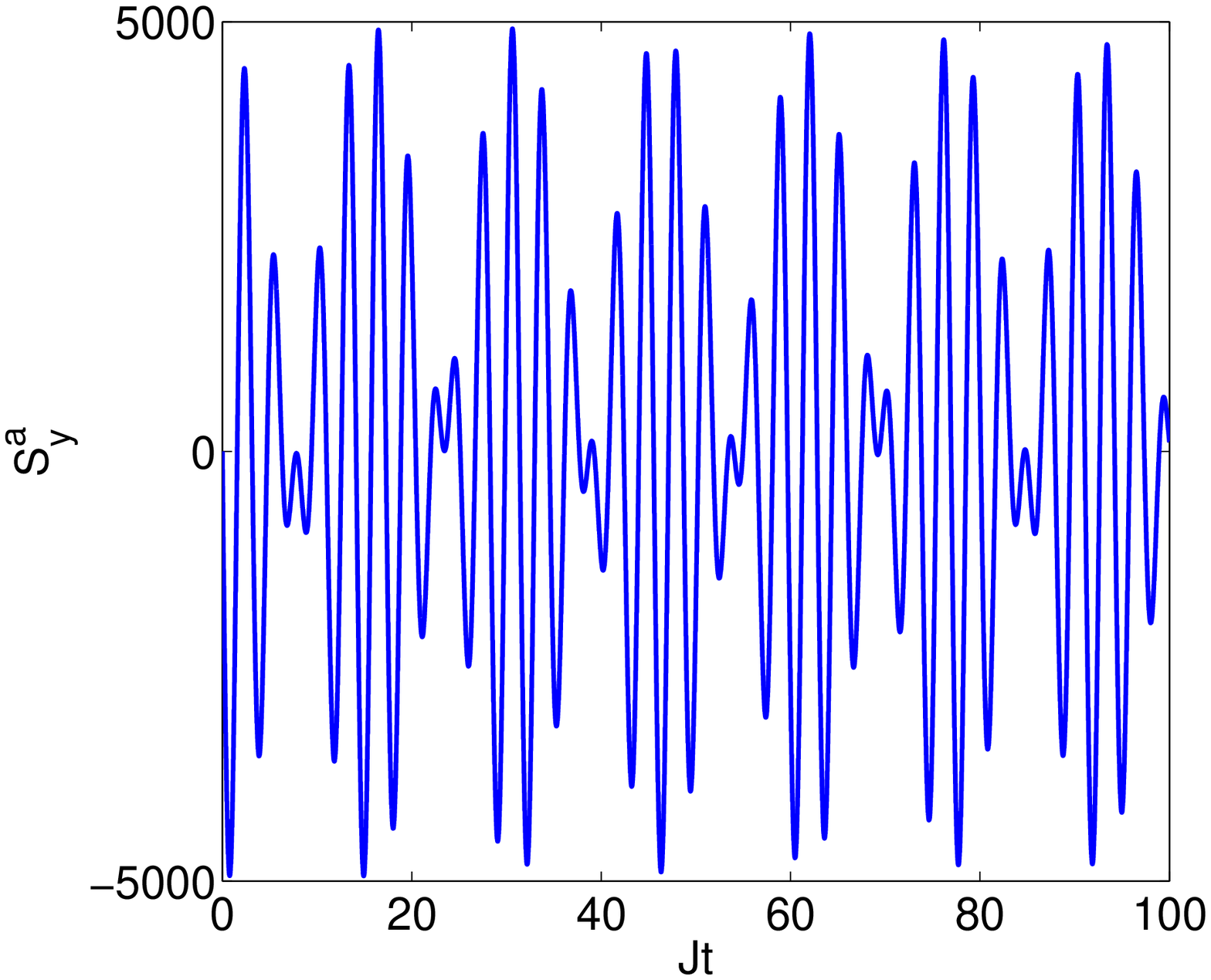}
\label{fig:Sya1f}}
\end{center}
\caption{(colour online) The longer time dynamics of the momenta between wells $a_{1}$ and $a_{2}$ for the same parameters and initial conditions as in Fig.~\ref{fig:Na1compare}. \subref{fig:Sya1c} shows the dynamics for initial coherent states, while \subref{fig:Sya1f} is for initial Fock states. The momenta between $b_{1}$ and $b_{2}$ are equal in magnitude and opposite in sign.}
\label{fig:intmomenta01}
\end{figure}

\begin{figure}
\begin{center}
\subfigure{\includegraphics[width=0.45\columnwidth]{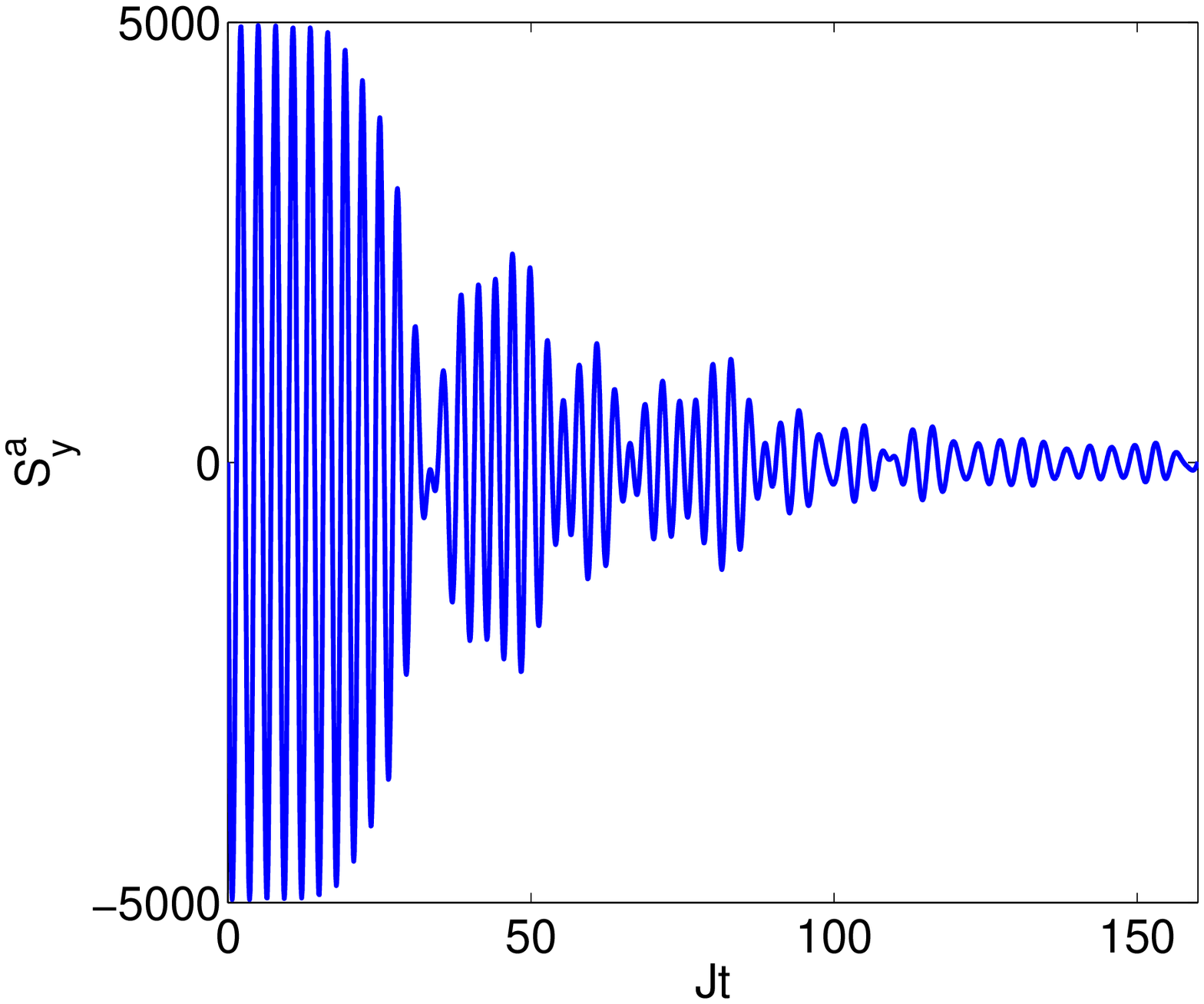}
\label{fig:Sya5c}}
\hspace{8pt}
\subfigure{\includegraphics[width=0.45\columnwidth]{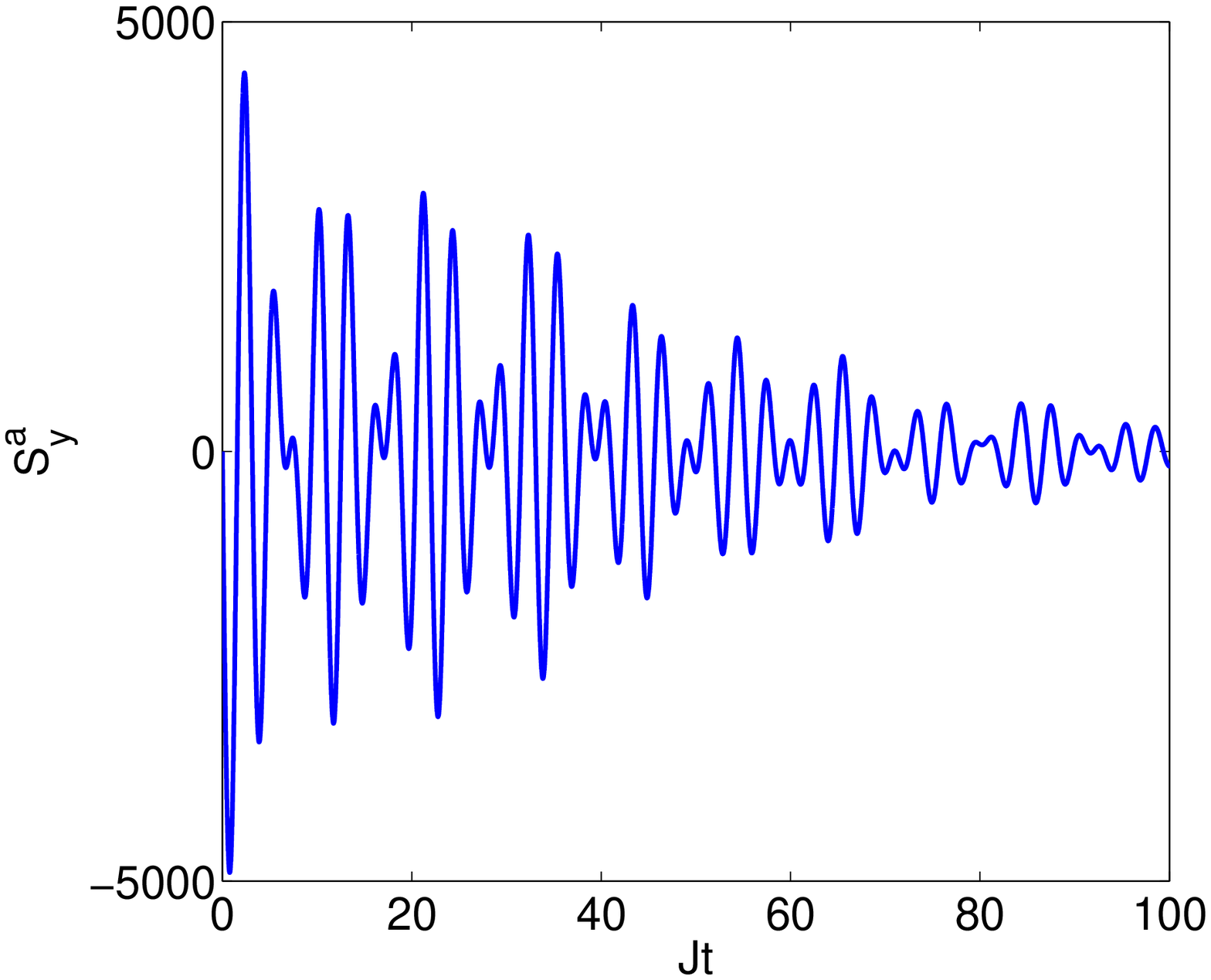}
\label{fig:Sya5f}}
\end{center}
\caption{(colour online) The longer time dynamics of the momenta between wells $a_{1}$ and $a_{2}$ for the same parameters and initial conditions as in Fig.~\ref{fig:Na1compare}, but with $\chi = 0.5J/N_{T}$ . \subref{fig:Sya5c} shows the dynamics for initial coherent states, while \subref{fig:Sya5f} is for initial Fock states. The momenta between $b_{1}$ and $b_{2}$ are equal in magnitude and opposite in sign.}
\label{fig:intmomenta05}
\end{figure}

Figs.~\ref{fig:intmomenta01} and \ref{fig:intmomenta05} show the momenta between the two wells within each subsystem, for two different collisional nonlinearities and initial quantum states. The differences between the two initial states here are markedly qualitative, with obviously different oscillation frequencies. The longer time behaviour suggests that in all four cases the system tends towards an equilibrium situation where the momenta will equal zero, but the envelopes under which this behaviour occurs are quite different. We found that the momenta between the two subsystems exhibits similar behaviours. While the coherent state collapses and revivals are naturally expected, we also see that decaying and not completely periodic behaviour is seen for initial Fock states.
This behaviour in particular suggests that there are a number of frequencies involved here with repeating and almost regular patterns being observable. This will be examined further in Section~\ref{sec:Josephson}, where we will compare the frequencies found in our numerical analysis with those predicted by Strzys and Anglin in their analytic approximation, and with those predicted by Bogoliubov theory.

\begin{figure}
\begin{center}
\subfigure{\includegraphics[width=0.45\columnwidth]{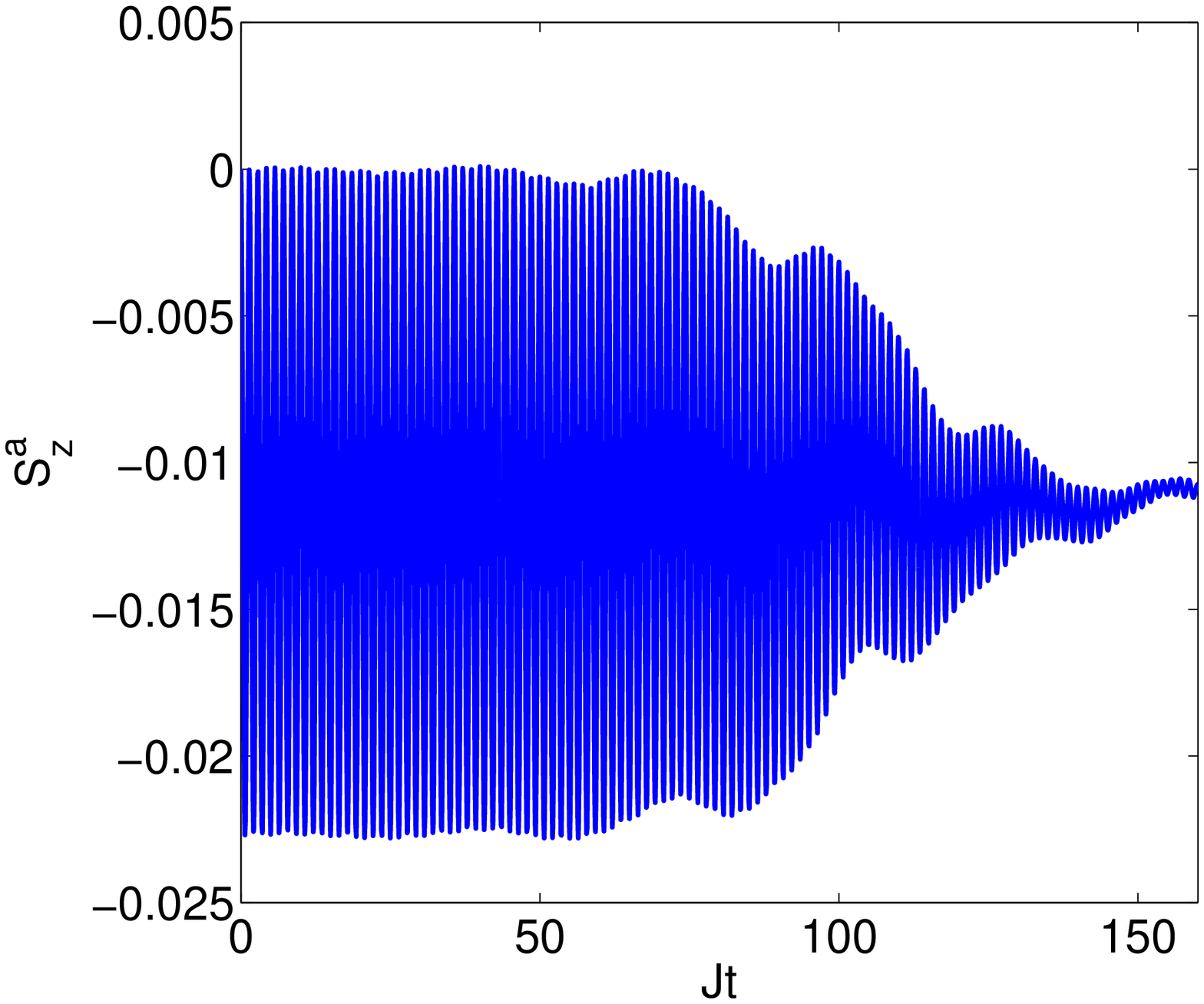}
\label{fig:Sza1c}}
\hspace{8pt}
\subfigure{\includegraphics[width=0.45\columnwidth]{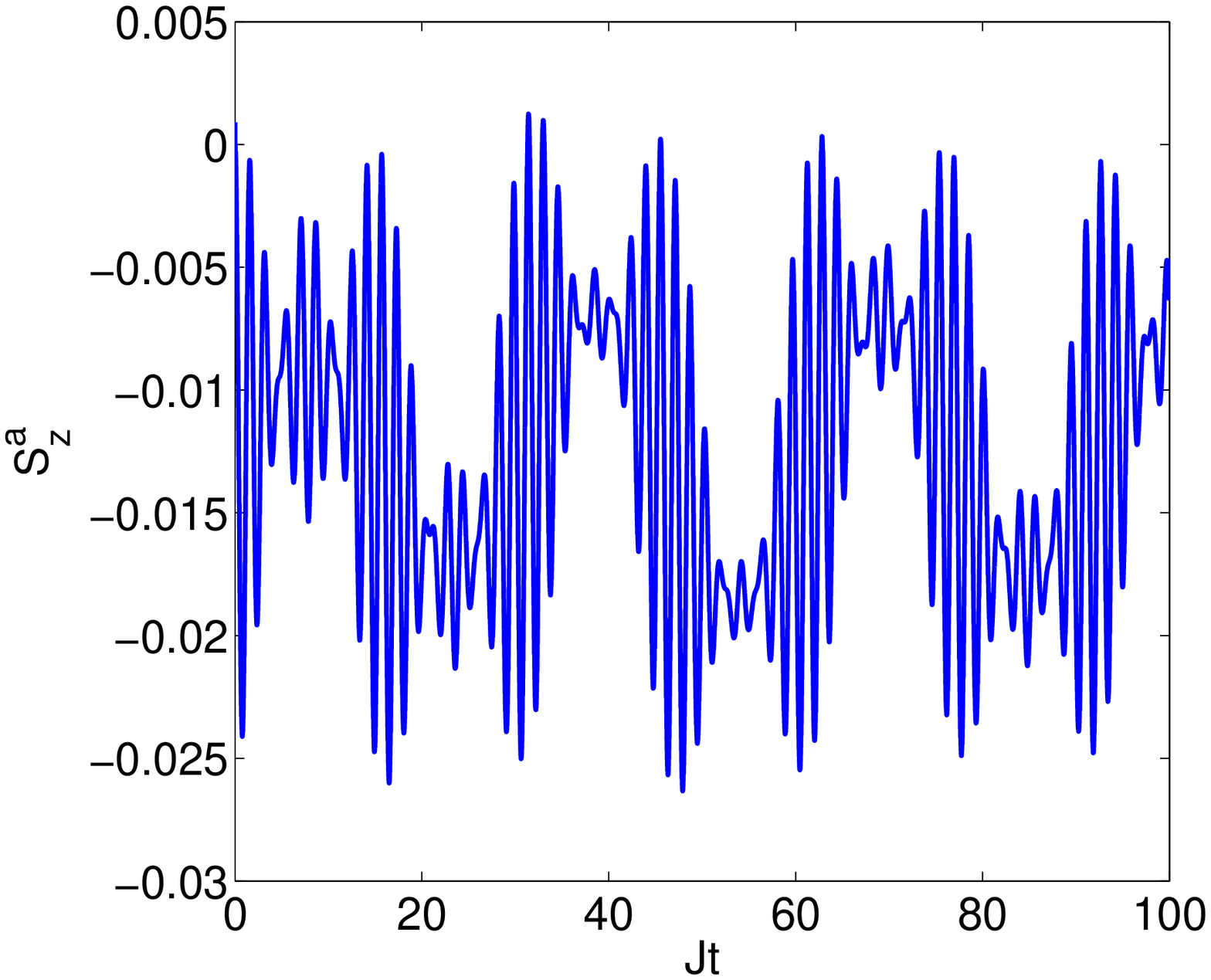}
\label{fig:Sza1f}}
\end{center}
\caption{(colour online) The longer time dynamics of the normalised $S_{z}^{a}$ for the same parameters and initial conditions as in Fig.~\ref{fig:Na1compare}. \subref{fig:Sza1c} shows the dynamics for initial coherent states, while \subref{fig:Sza1f} is for initial Fock states.}
\label{fig:superfluida1}
\end{figure}

\begin{figure}
\begin{center}
\subfigure{\includegraphics[width=0.45\columnwidth]{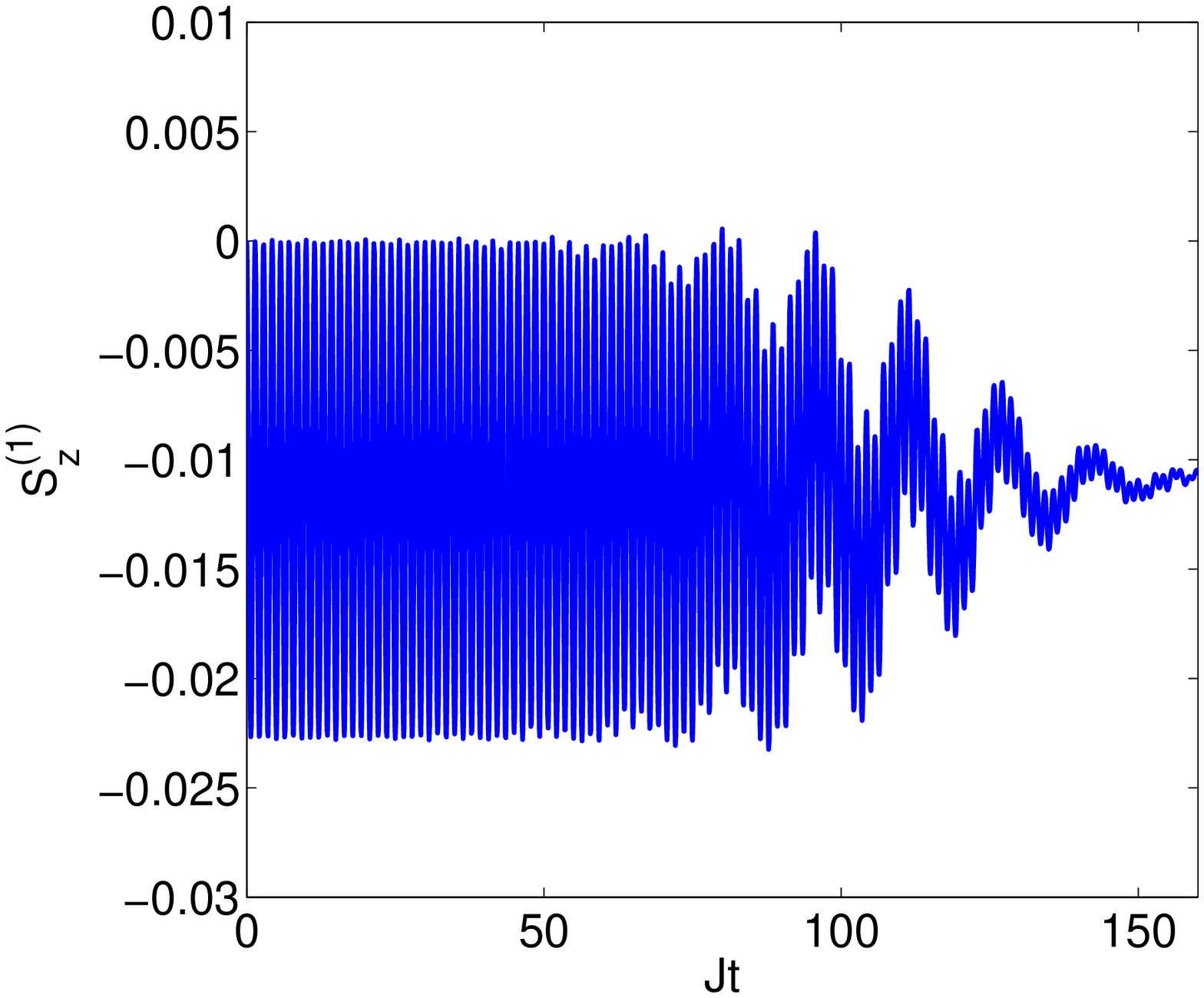}
\label{fig:Sz11c}}
\hspace{8pt}
\subfigure{\includegraphics[width=0.45\columnwidth]{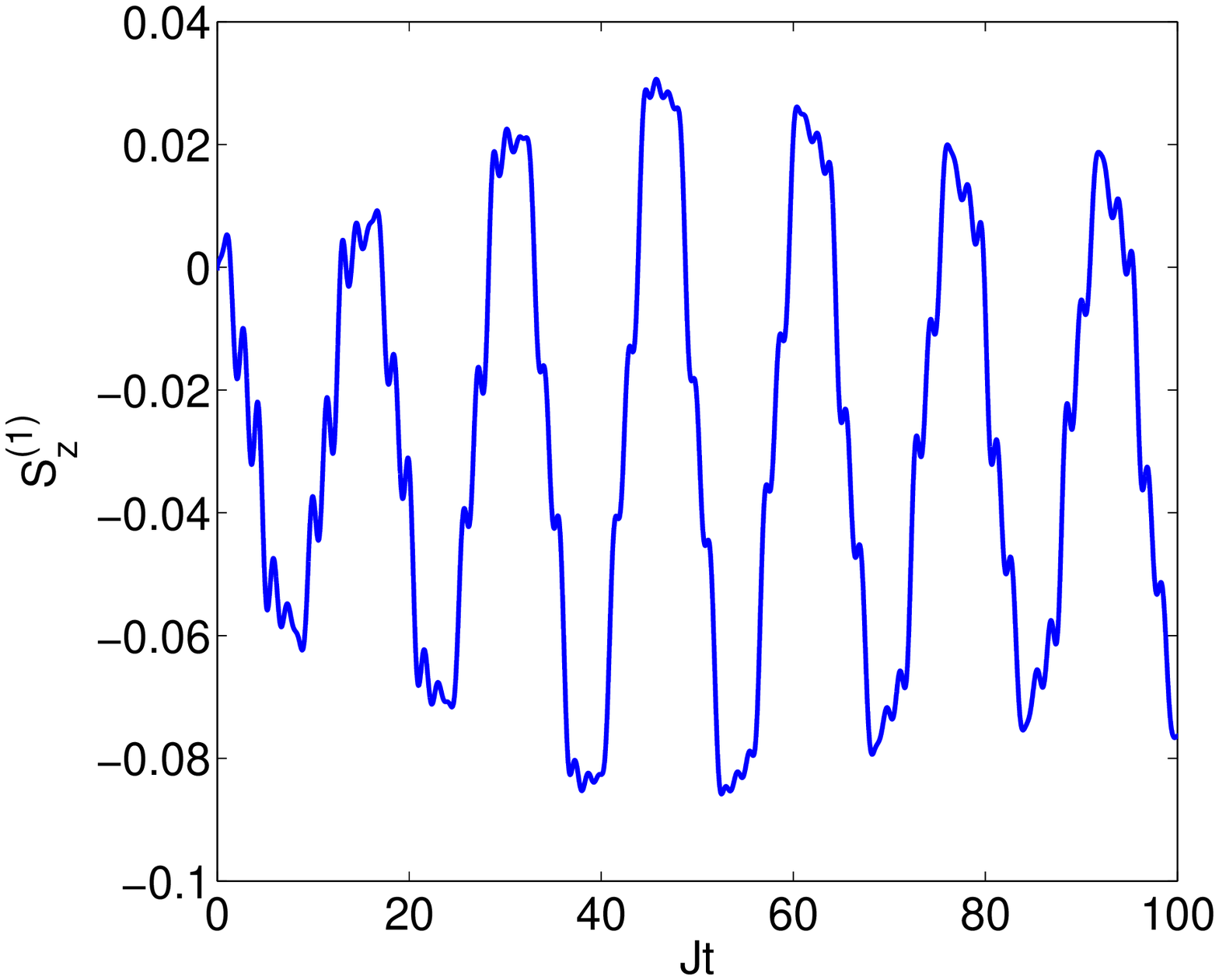}
\label{fig:Sz11f}}
\end{center}
\caption{(colour online) The longer time dynamics of the normalised $S_{z}^{(1)}$ for the same parameters and initial conditions as in Fig.~\ref{fig:Na1compare}. \subref{fig:Sz11c} shows the dynamics for initial coherent states, while \subref{fig:Sz11f} is for initial Fock states.}
\label{fig:superfluidab1}
\end{figure}

\begin{figure}
\begin{center}
\subfigure{\includegraphics[width=0.45\columnwidth]{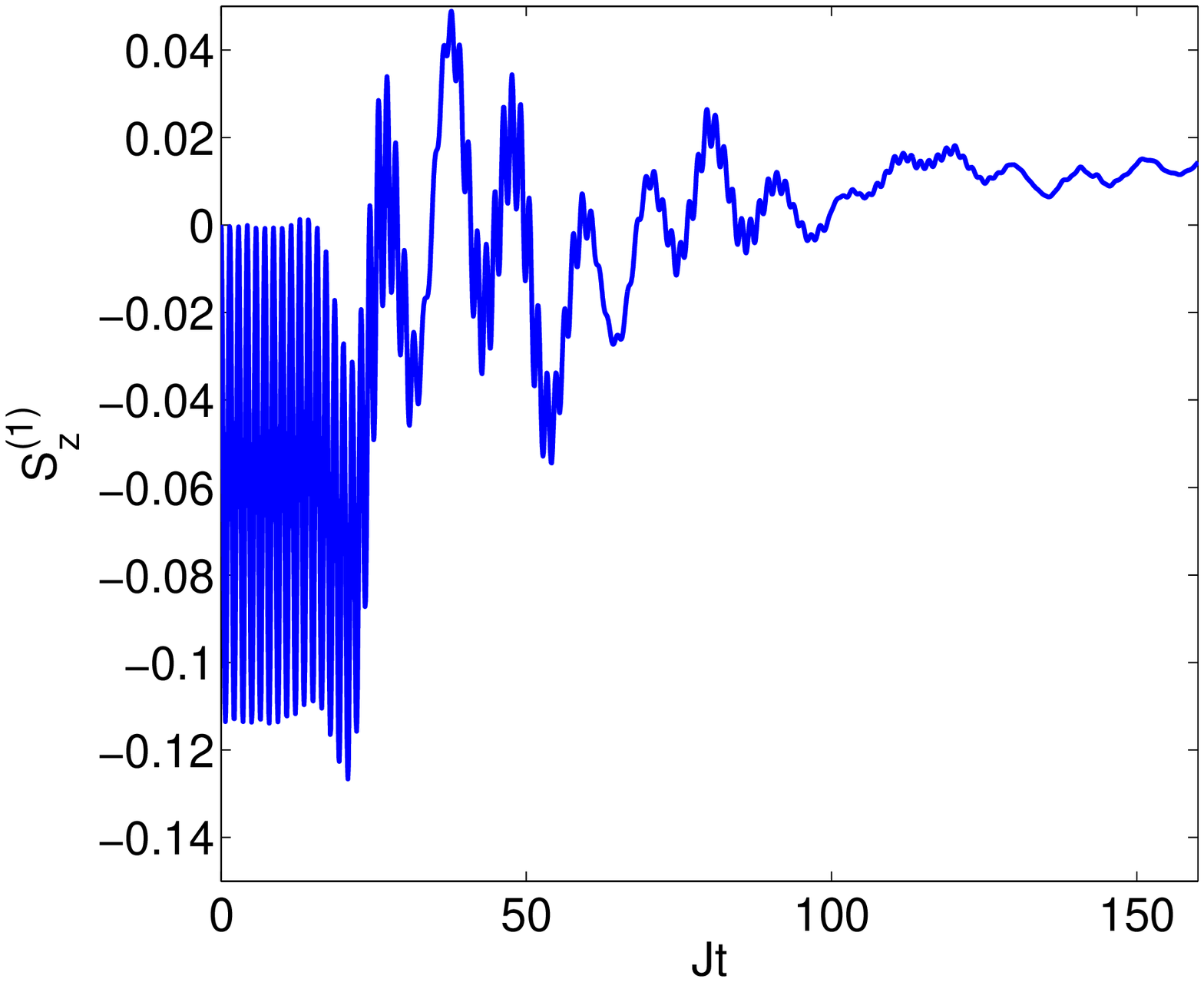}
\label{fig:Sz15c}}
\hspace{8pt}
\subfigure{\includegraphics[width=0.45\columnwidth]{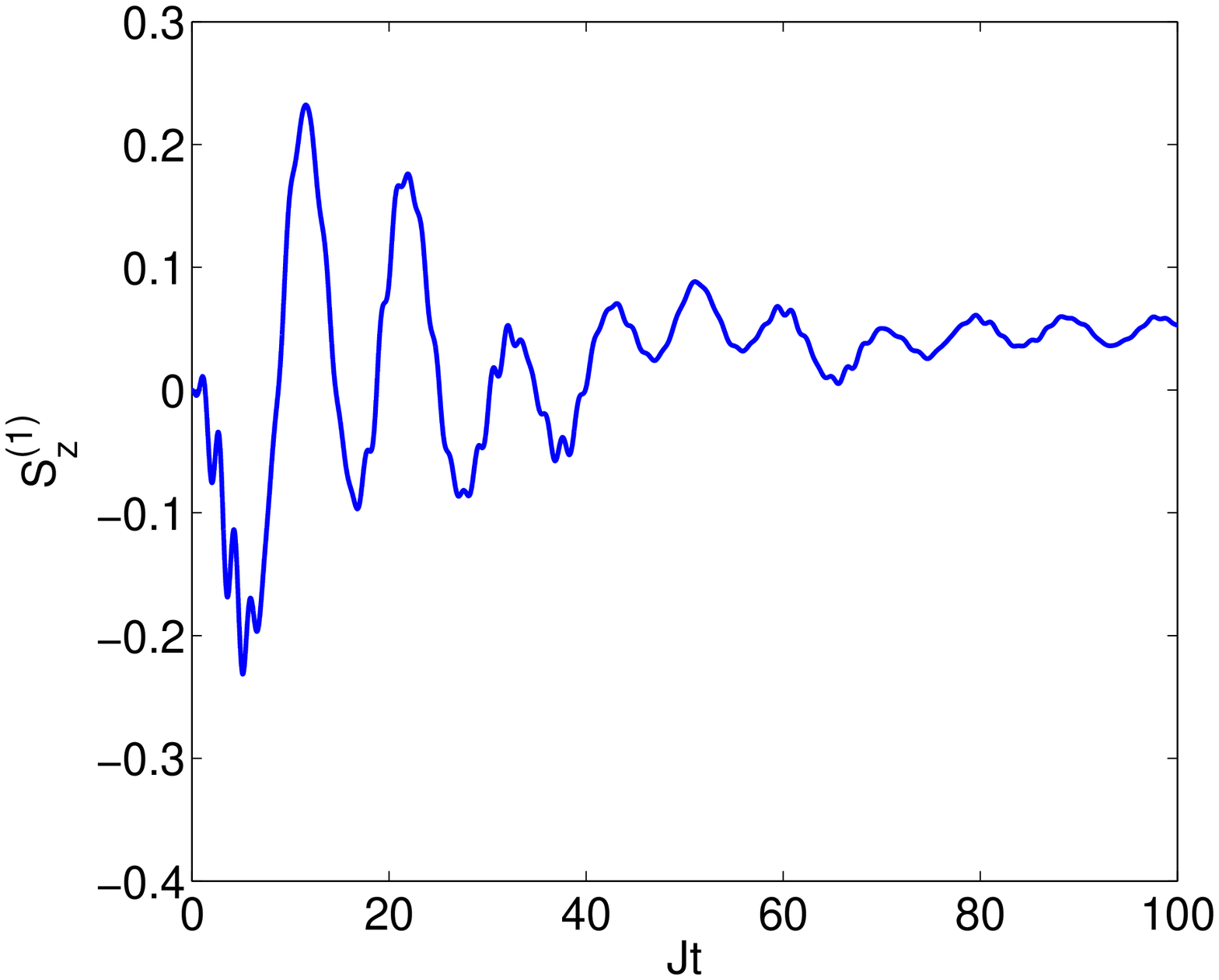}
\label{fig:Sz15f}}
\end{center}
\caption{(colour online) The longer time dynamics of the normalised $S_{z}^{(1)}$ for the same parameters and initial conditions as in Fig.~\ref{fig:Na1compare}, but with $\chi = 0.5J/N_{T}$. \subref{fig:Sz15c} shows the dynamics for initial coherent states, while \subref{fig:Sz15f} is for initial Fock states.}
\label{fig:superfluidab5}
\end{figure}

In Fig.~\ref{fig:superfluida1} we show $S_{z}$ between wells $a_{1}$ and $a_{2}$ for the lower collisional nonlinearity, finding once again that the dynamics differ qualitatively for the different initial quantum states. In the case of initial coherent states, the oscillations in $S_{z}^{a}$ decay to a finite value, while the Fock state case continues to oscillate over the time shown here in a complex, but almost periodic manner. We see the same type of behaviour for $S_{z}^{(1)}$, which is between the two halves of the system, as shown in Fig.~\ref{fig:superfluidab1} and Fig.~\ref{fig:superfluidab5}. What is obvious here is that, when the interaction strength is increased, $S_{z}$ and all the other quantities we have calculated settle down faster to a type of equilibrium situation where the atoms are evenly distributed among the four wells and the momenta and tunnelling energies are not changing very much. In fact, we would only expect this system to oscillate indefinitely for a vanishing atomic interaction strength, in which case it would be composed of ideal gases and the semiclassical analysis would be accurate. 
The difference in these behaviours is not predictable from the analysis used by Strzys and Anglin, but will be seen to be important when we investigate the analogies with heat exchange and entropy, below in section~\ref{sec:calor}. 

\section{Josephson Frequencies}
\label{sec:Josephson}

Now that we have shown that the classical solutions for the dynamics of this system are not always reliable, we will examine our results for evidence of the Josephson oscillations predicted in the linear Bogoliubov approximation, as well as the low frequency collective mode predicted by Strzys and Anglin. In terms of the units we use, the Josephson frequencies for elementary excitations above the N-atom ground state are
\begin{eqnarray}
\tilde{\omega} &=& \sqrt{2\omega(2\omega+N_{T}\chi)},\nonumber\\
\tilde{\Omega} &=& \sqrt{2J(2J+N_{T}\chi)},\nonumber\\
\tilde{\Omega}^{\prime} &=& \sqrt{2(J+\omega)(2J+2\omega+N_{T}\chi)}.
\label{eq:Josfreq}
\end{eqnarray}
In this case, $\tilde{\Omega}^{\prime}$ and $\tilde{\Omega}$ are at a higher frequency than $\tilde{\omega}$ and Stryzs and Anglin also predict a beating between these two, with frequency
\begin{eqnarray}
\omega_{J} &=&  \tilde{\Omega}^{\prime}-\tilde{\Omega}\nonumber\\
& \approx &   2 \omega+\frac{\omega(2\omega+N_{T}\chi)}{4J},
\label{eq:beating}
\end{eqnarray}
when $J\gg\omega$. In the cases we examine in this article, $J=1$, while $\omega=0.1$ and $N_{T}\chi$ has so far been either $0.1$ or $0.5$, so this condition holds.

\begin{figure}
\begin{center}
\subfigure{\includegraphics[width=0.45\columnwidth]{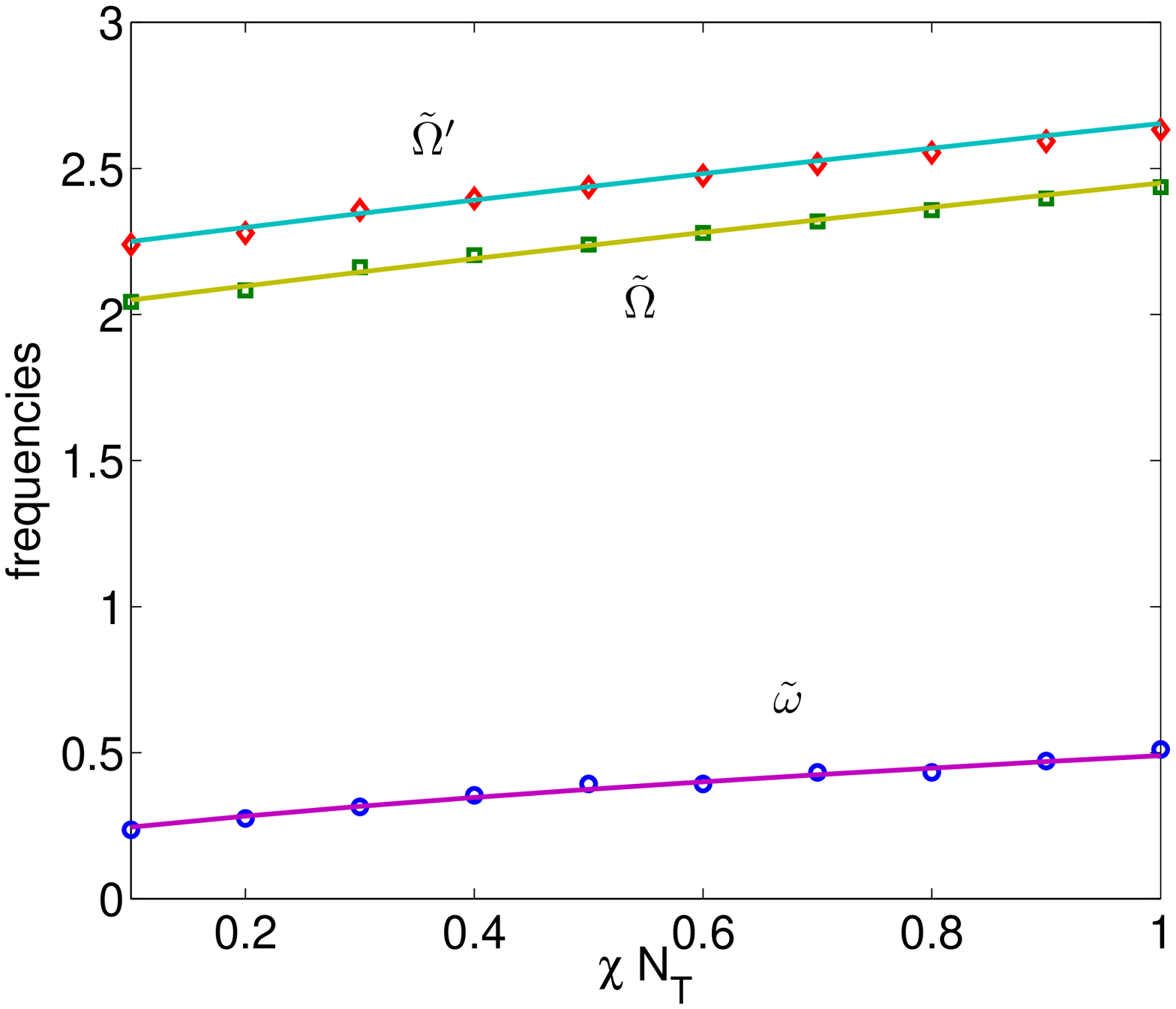}
\label{fig:FreqCohA}}
\hspace{8pt}
\subfigure{\includegraphics[width=0.45\columnwidth]{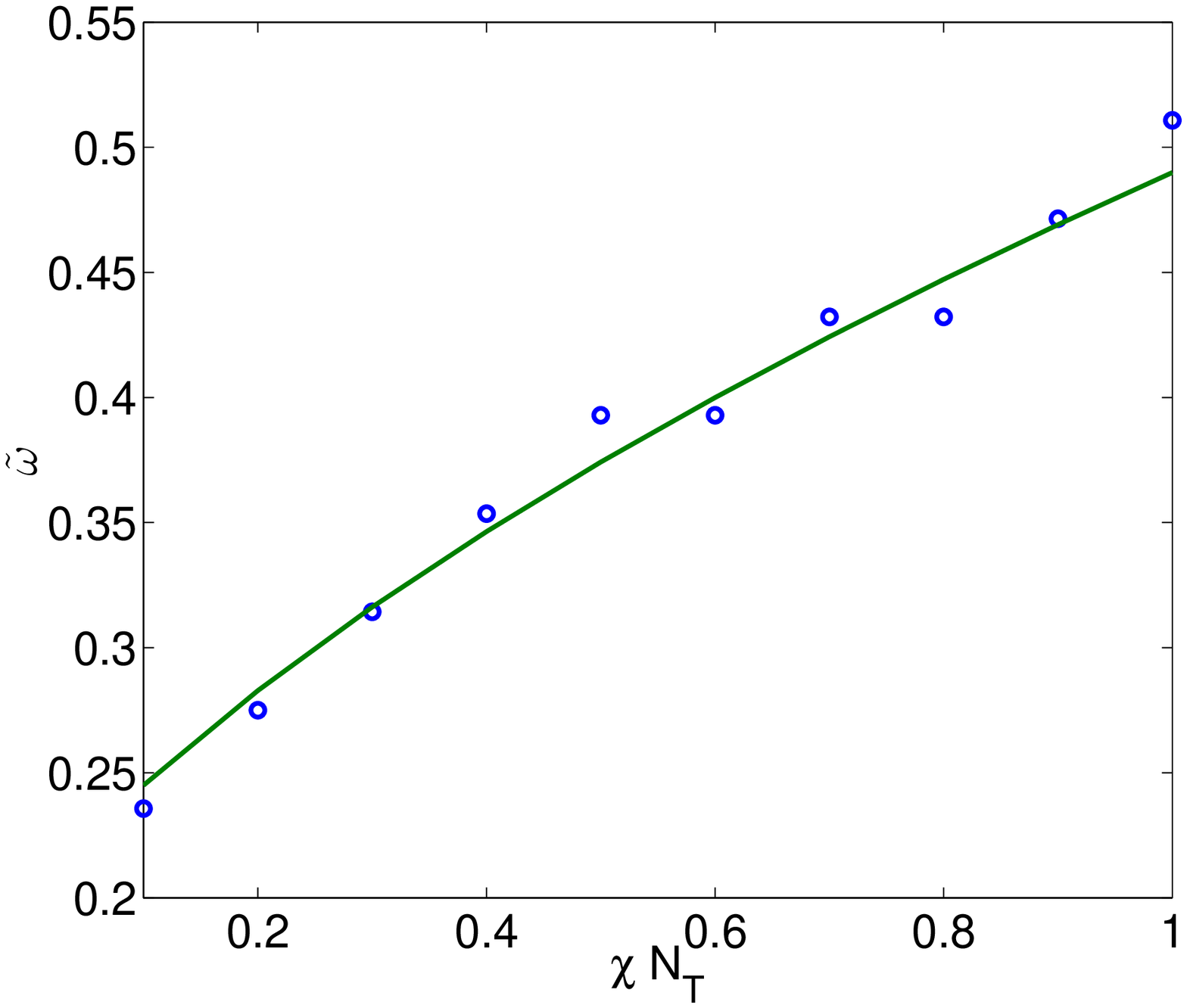}  
\label{fig:FreqCohB}}
\end{center}
\caption{(colour online) The frequencies of the Josephson oscillations found numerically for initial coherent states, with 
   the solid lines being the analytical expressions. \subref{fig:FreqCohA} shows the results  for an equal initial distribution of atoms in each well, i.e. $N_{a_{1}}(0)=N_{a_{2}}(0)=N_{b_{1}}(0)=N_{b_{2}}(0)=2,500$, while \subref{fig:FreqCohB} initially has $N_{a_{1}}(0)=N_{a_{2}}(0)=2600$ and $N_{b_{1}}(0)=N_{b_{2}}(0)=2,400$ .
  The tunneling interaction strengths are $J = 1$ and $\omega = 0.1$. }
\label{fig:Freqs}
\end{figure}

We now look for evidence of these frequencies in Fourier transforms of the expectation values of the atomic numbers and the $S$ operators, using different initial quantum states and number distributions. In Fig.~\ref{fig:FreqCohA}, we begin with the populations equally distributed among the wells in coherent states and take the Fourier transforms of the number in any one well. As can clearly be seen, the observed frequencies follow the Bogoliubov predictions very closely for this configuration. We note here that the actual number oscillations are very small, being driven by the Poissonian number uncertainties in each well.  
In this case we do not see clear evidence in the Fourier components of the beat mode predicted by Strzys and Anglin, although we note that it would be only marginally different from $\tilde{\omega}$. In view of the fact that Strzys and Anglin found their numerical results by adding a time dependent potential tilt to each subsystem, we also began with an asymmetric system, with $2600$ atoms in each of the wells on the left hand side, and $2400$ in each of the others. As shown in Fig.~\ref{fig:FreqCohB}, this resulted in one frequency of oscillation only for the atoms, with the Fourier transforms being extremely clean, and this frequency closely matching $\tilde{\omega}$. We have not included frequency analysis of the system beginning in initial Fock states, as we found that these needed averaging over too many stochastic trajectories to give clear signals.

\begin{figure}
\begin{center}
\subfigure{\includegraphics[width=0.45\columnwidth]{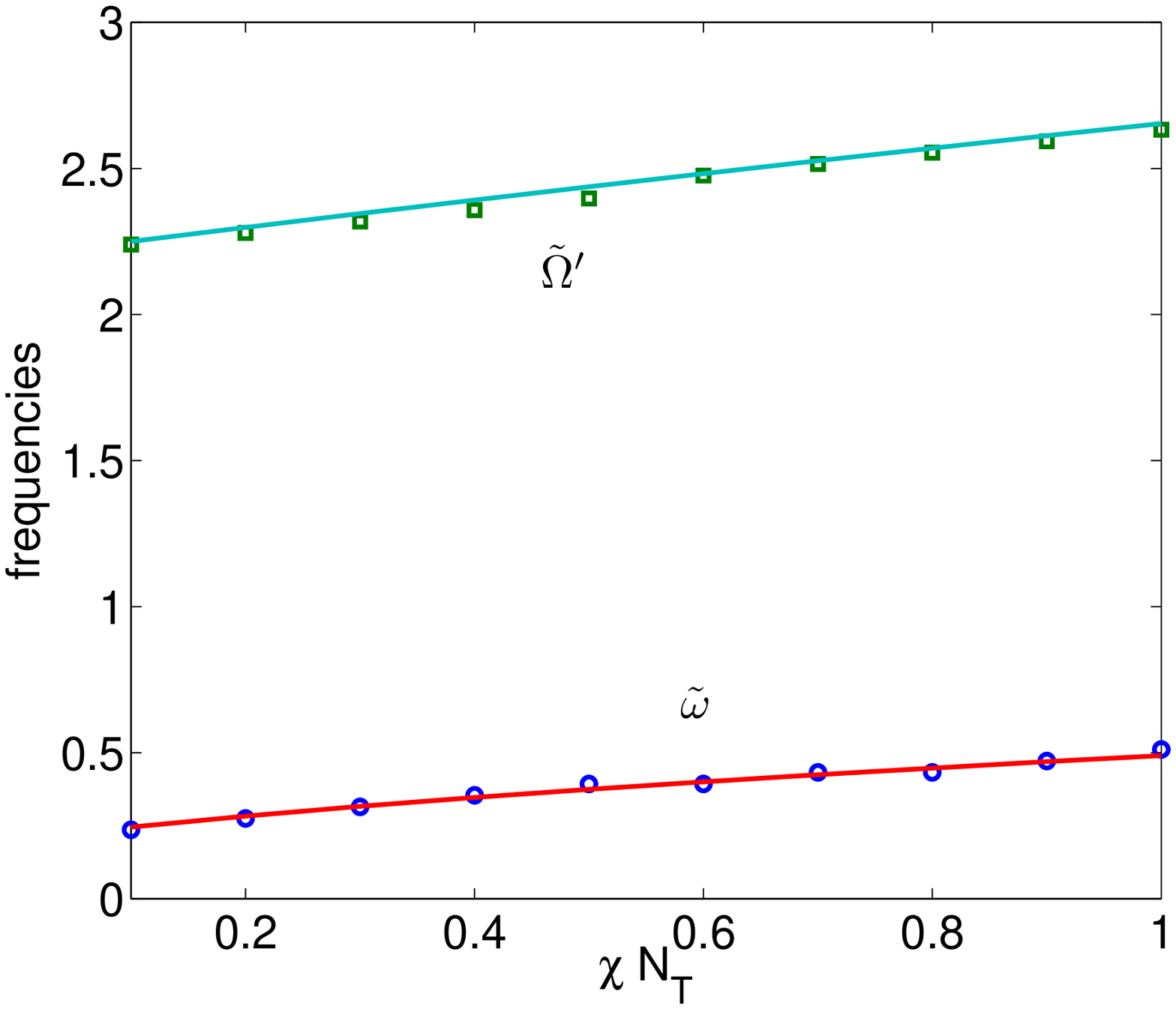}
\label{fig:Sy1even}}
\hspace{8pt}
\subfigure{\includegraphics[width=0.45\columnwidth]{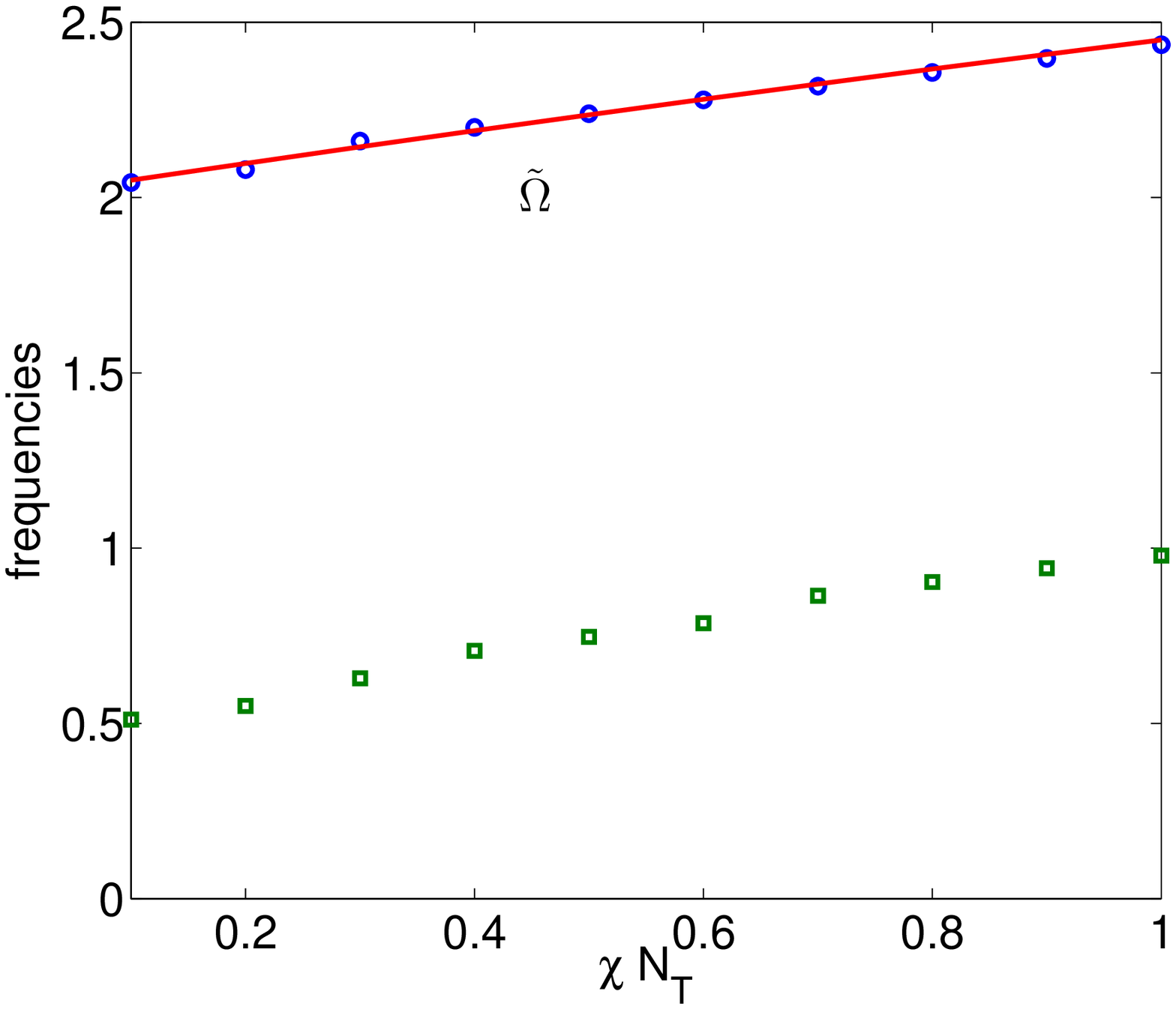}  
\label{fig:Sz1even}}
\end{center}
\caption{(colour online) The frequencies of the oscillations found numerically for initial coherent states from $S_{y}^{(1)}$ and $S_{z}^{(1)}$, with 
   the solid lines being the analytical expressions. The initial atom numbers are evenly distributed. \subref{fig:Sy1even} shows the results from $S_{y}^{(1)}$, while \subref{fig:Sz1even} is for $S_{z}^{(1)}$ .
  The tunneling interaction strengths are $J = 1$ and $\omega = 0.1$. }
\label{fig:Szyeven}
\end{figure}

\begin{figure}
\begin{center}
\subfigure{\includegraphics[width=0.45\columnwidth]{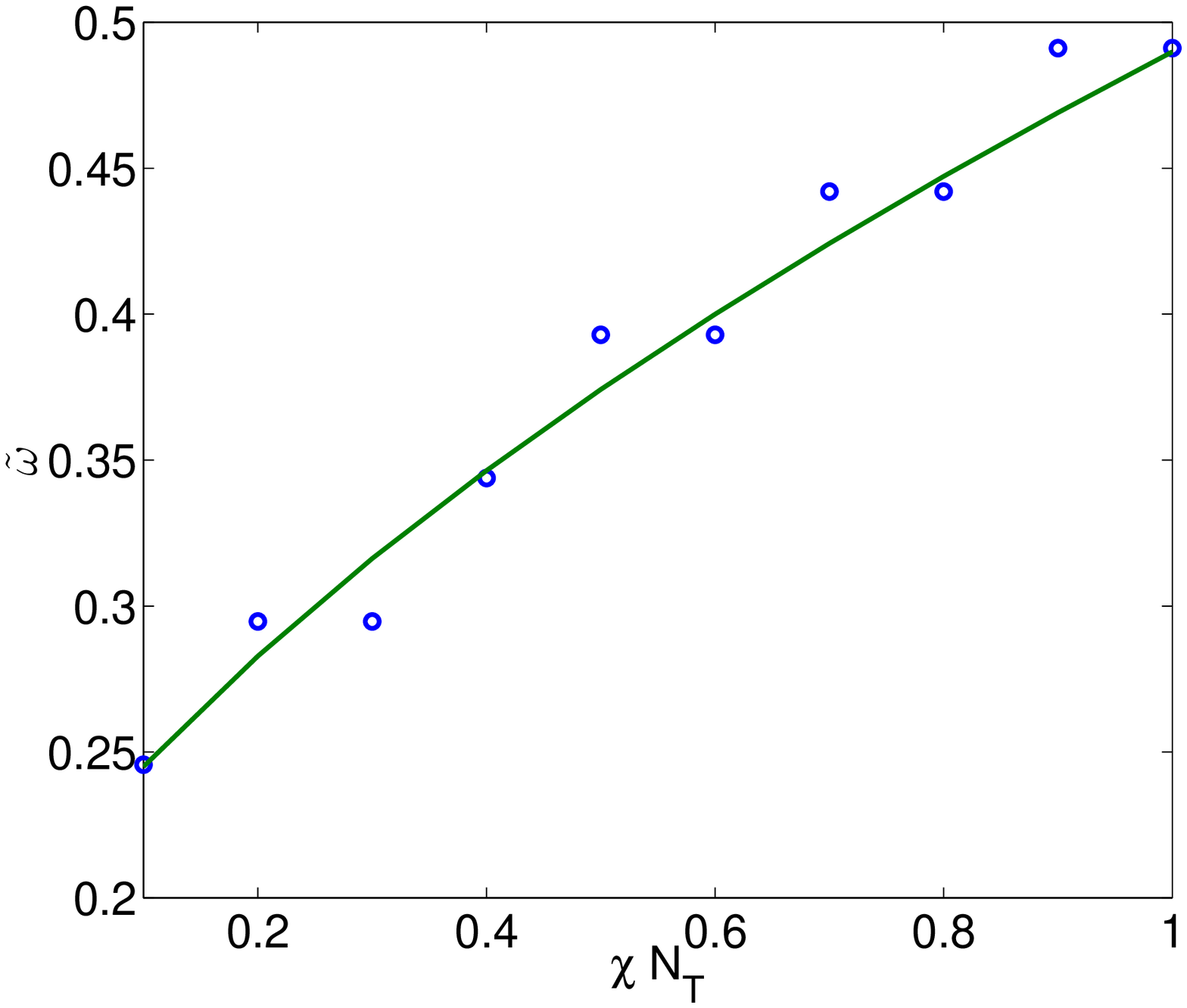}
\label{fig:Sy1odd}}
\hspace{8pt}
\subfigure{\includegraphics[width=0.45\columnwidth]{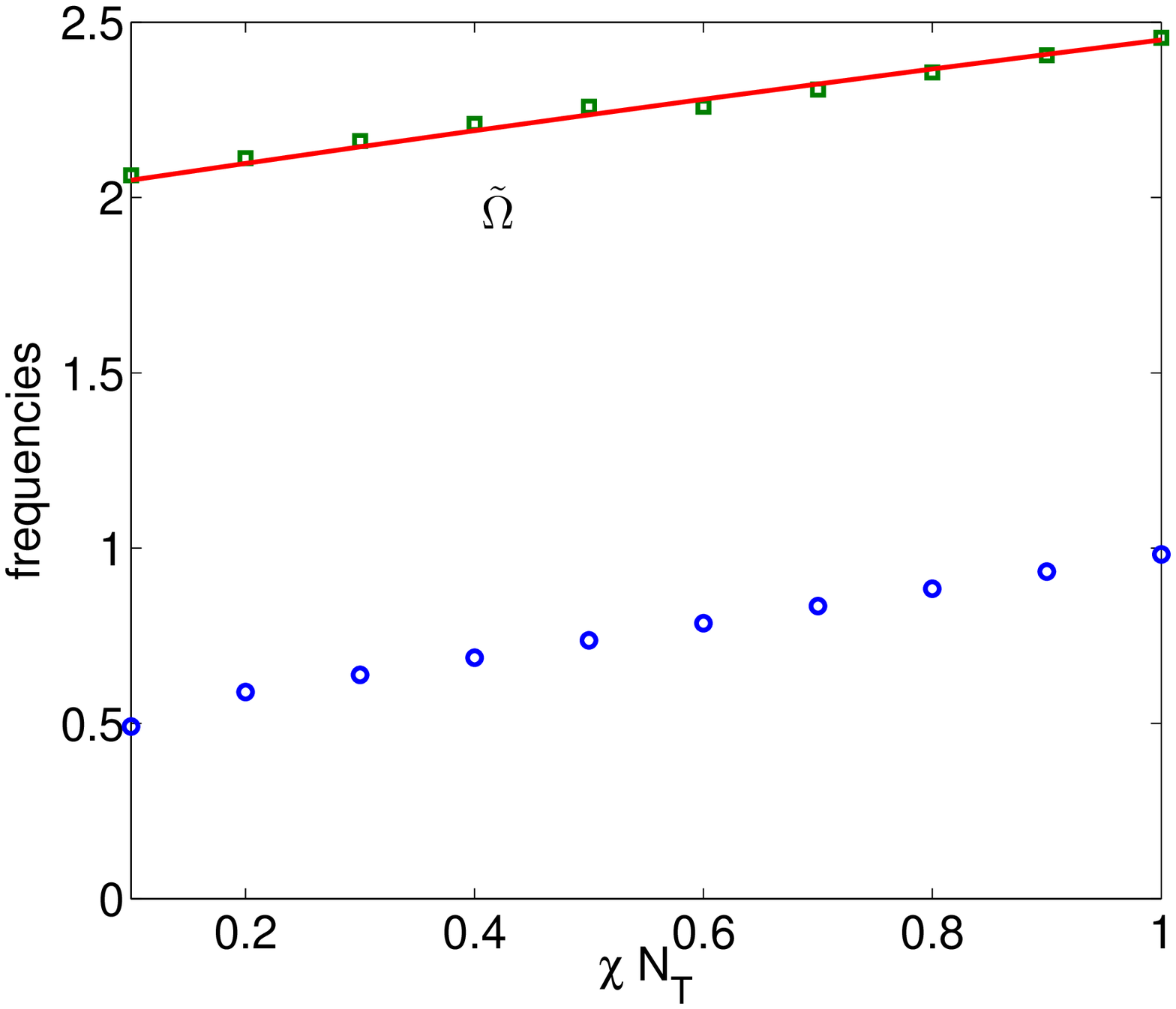}  
\label{fig:Sz1odd}}
\end{center}
\caption{(colour online) The frequencies of the oscillations found numerically for initial coherent states from $S_{y}^{(1)}$ and $S_{z}^{(1)}$, with 
   the solid lines being the analytical expressions. The initial atom numbers are $N_{a_{1}}(0)=N_{a_{2}}(0)=2600$ and $N_{b_{1}}(0)=N_{b_{2}}(0)=2,400$. \subref{fig:Sy1odd} shows the results from $S_{y}^{(1)}$, while \subref{fig:Sz1odd} is for $S_{z}^{(1)}$ .
  The tunneling interaction strengths are $J = 1$ and $\omega = 0.1$. }
\label{fig:Szyodd}
\end{figure}

We also examined the frequencies of oscillation in $S_{y}^{(1)}$ and $S_{z}^{(1)}$, as these represent tunnelling momentum and interaction between the two subsytems. These are across the weak link where Strzys and Anglin expect a process analogous to heat transfer to take place, so it is of interest to look for evidence of the collective mode frequencies predicted in their paper. When we examine the results for an initial equal number distribution, we find frequencies closely matching $\tilde{\omega}$ and $\tilde{\Omega}^{\prime}$ in $S_{y}^{(1)}$, and two clear frequencies in $S_{z}^{(1)}$, one of which matches $\tilde{\Omega}$, while the other is not close to any of the four frequencies we have dealt with so far. As far as we can tell, it does not seem to be either of $\tilde{\omega}_{\pm}$ from Eq.~(12) of the Strzys and Anglin paper, as these should be either comparable to or lower than $\tilde{\omega}$, although it is difficult to deconstruct their approximations to know what the joson number should be for our parameters. These results are shown in Fig.~\ref{fig:Szyeven}.
The frequencies found by beginning with the unequal number distribution are shown in Fig.~\ref{fig:Szyodd}, where we again see that not all the analytically predicted results are clearly found. We note here that we did not attempt to mechanically excite these predicted frequencies, but instead looked for those which may arise naturally from population imbalances in the system. 

\section{Analogy with Josephson heat oscillations}
\label{sec:calor}

\begin{figure}
\begin{center}
\subfigure{\includegraphics[width=0.45\columnwidth]{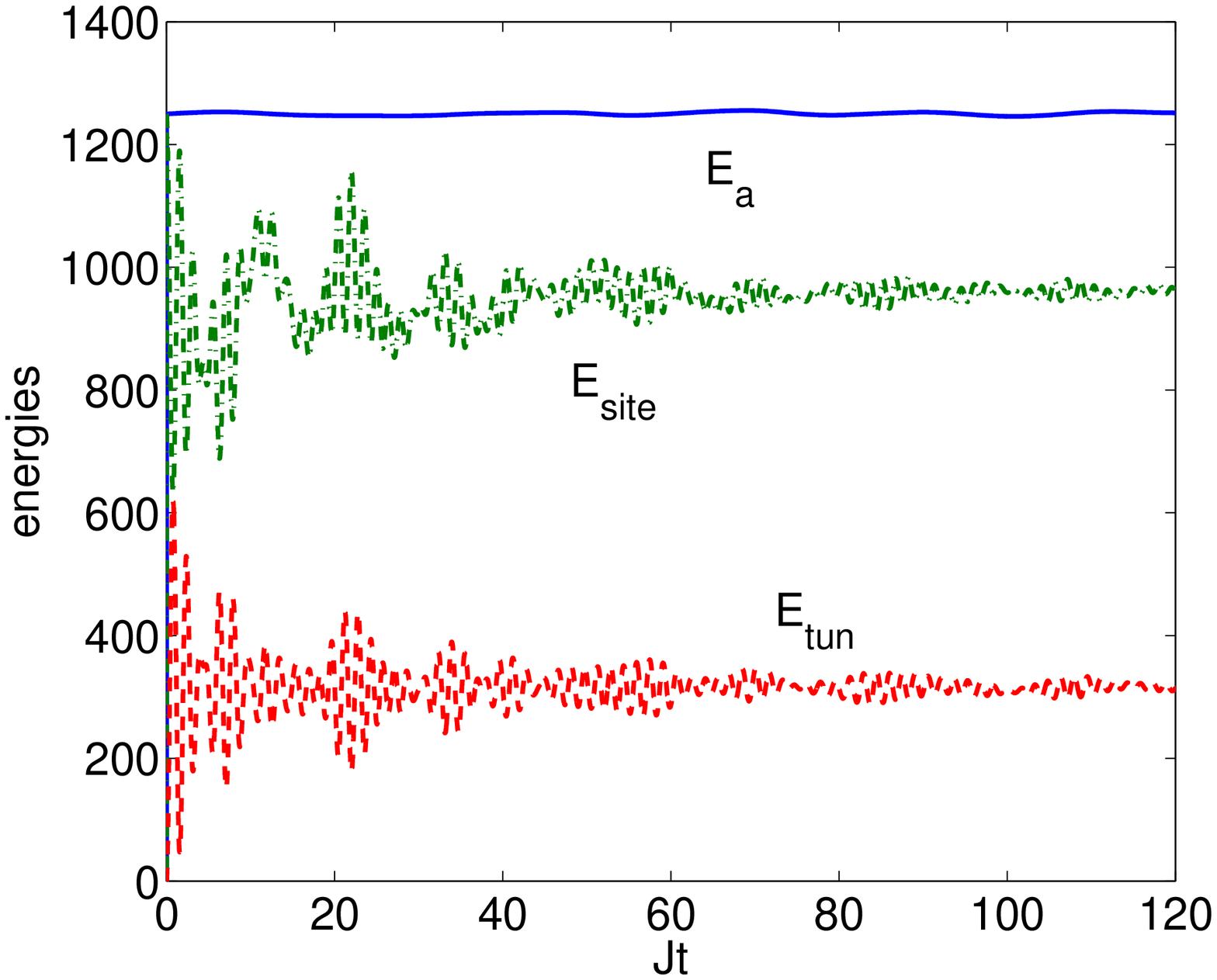}
\label{fig:energies}}
\hspace{8pt}
\subfigure{\includegraphics[width=0.45\columnwidth]{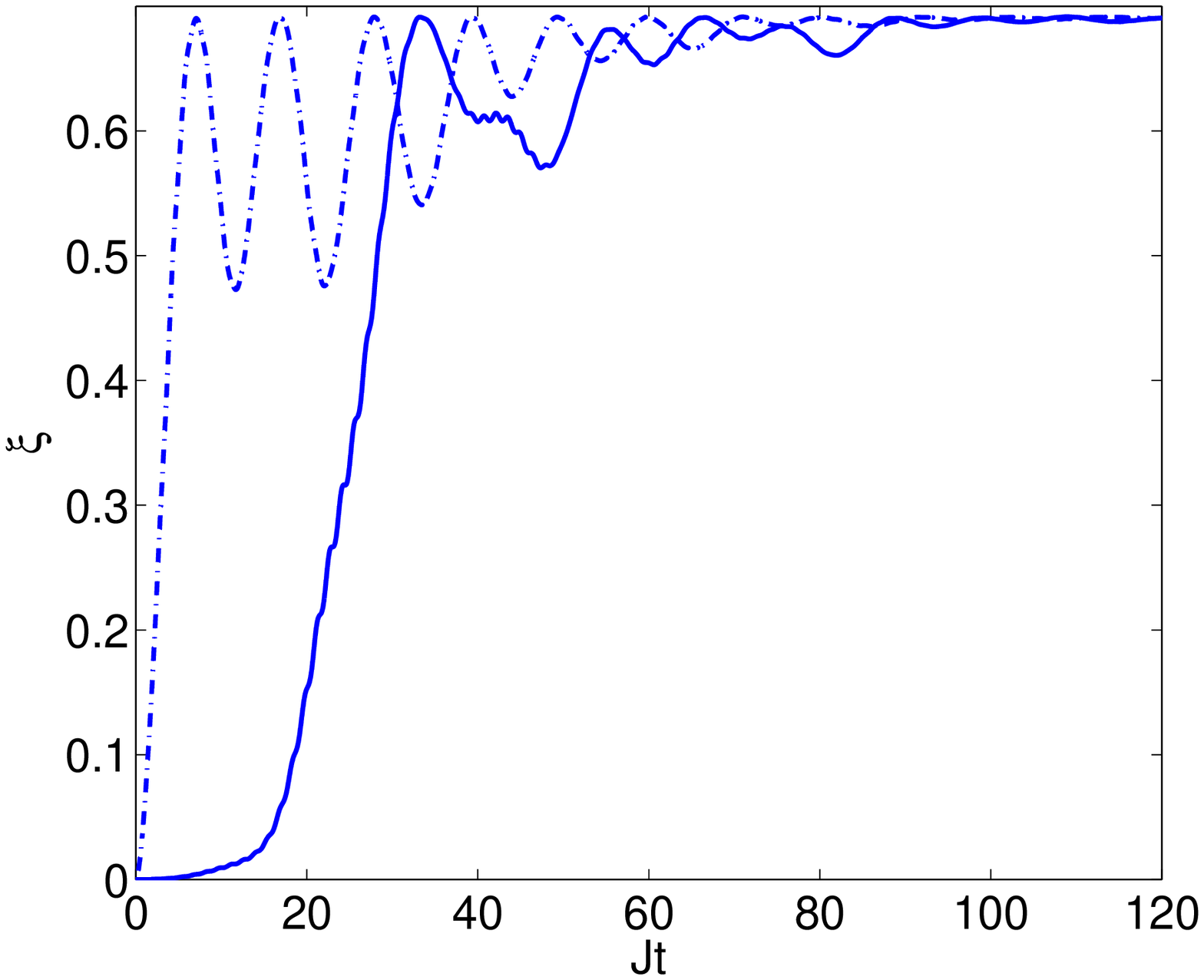}  
\label{fig:entropy}}
\end{center}
\caption{(colour online) \subref{fig:energies} The different energies in the left subsystem as a function of scaled time, for initial Fock states with $N_{a_{1}}(0)=N_{b_{2}}(0)=5000$ and $N_{a_{2}}(0)=N_{b_{1}}(0)=0$, $J=1$, $\omega=0.1$ and $\chi N_{T}=0.5$. \subref{fig:entropy} The pseudo entropy for the same parameters. The solid line is for initial coherent states while the dash-dotted line is for initial Fock states.}
\label{fig:energyandentropy}
\end{figure}

As the principal motivation behind the work of Strzys and Angin was as a contribution to the development of a mesoscopic and eventually microscopic model for heat transfer, we will now examine the quantities in our system which may be useful in such a model. Heat transfer involves two principal processes. The first of these is energy and the second is a change in entropy if we begin in a non-equilibrium state, as we do in this article. The expectation values of the various energies involved can be calculated straightforwardly by finding the dynamical averages of the quantities in the effective Hamiltonian of Eq.~(\ref{eq:effhamiltonian}).  As expected for a closed system, the total energy is a constant. What is interesting, however, is that even as the mean values of the oscillations die down and the atoms approach a state where they are evenly distributed among the four wells, the tunnelling energy between sites does not disappear, but approaches a constant value. This is shown in Fig.~\ref{fig:energies}, where we have plotted the different components for the left hand side of the system, where the different energies shown are defined as (note that we set $\hbar=1$ for the graphics so that we are not using S.I. units)
\begin{eqnarray}
E_{site} &=& \hbar\chi\left(\hat{a}_{1}^{\dag\;2}\hat{a}_{1}^{2}+\hat{a}_{2}^{\dag\;2}\hat{a}_{2}^{2}\right),\nonumber\\
E_{tun} &=& -\hbar J \left(\hat{a}_{1}^{\dag}\hat{a}_{2}+\hat{a}_{2}^{\dag}\hat{a}_{1}\right),\nonumber\\
E_{a} &=& E_{site}+E_{tun}-\frac{\hbar\omega}{2}\left(\hat{a}_{1}^\dagger \hat{b}_{1}+\hat{b}_{1}^\dagger \hat{a}_{1}+\hat{a}_{2}^\dagger \hat{b}_{2}+\hat{b}_{2}^\dagger \hat{a}_{2}\right),
\label{eq:AsideE}
\end{eqnarray}
where half the weak link tunnelling energy has been included in the total energy of the left hand side. 

We can also calculate a candidate for entropy by following the approach taken by Strzys and Anglin in Eq.~(13) of their paper. Before we do this, however, we will examine their expression for the single-particle reduced density matrix in more detail. Considering only one of the subsystems, we may define the even and odd modes excited by $\hat{a}_{\pm}=(\hat{a}_{1}\pm\hat{a}_{2})$, (note that our definition differs by a scale factor of $1/\sqrt{2}$) which leads to the matrix,
\begin{eqnarray}
R_{\alpha}&=&\frac{\langle \hat{a}_{\pm}^{\dag}\hat{a}_{\pm} \rangle}{\langle \hat{a}_{1}^{\dag}\hat{a}_{1}+\hat{a}_{2}^{\dag}\hat{a}_{2}\rangle},\nonumber\\
&=& \frac{1}{\langle \hat{a}_{1}^{\dag}\hat{a}_{1}+\hat{a}_{2}^{\dag}\hat{a}_{2}\rangle}
\left(\begin{array}{cc}
\langle\hat{a}_{+}^{\dag}\hat{a}_{+}\rangle & \langle\hat{a}_{+}^{\dag}\hat{a}_{-}\rangle\\
\langle\hat{a}_{-}^{\dag}\hat{a}_{+}\rangle & \langle\hat{a}_{-}^{\dag}\hat{a}_{-}\rangle\end{array}\right),
\label{eq:aglinmat}
\end{eqnarray}
which is presented in ref.~\cite{Anglin} as a diagonal matrix. However, when we expand their definition, we find 
\begin{equation}
R_{\alpha} = \frac{1}{2\langle \hat{a}_{1}^{\dag}\hat{a}_{1}+\hat{a}_{2}^{\dag}\hat{a}_{2}\rangle}
\left(\begin{array}{cc}
\langle\hat{a}_{1}^{\dag}\hat{a}_{1}+\hat{a}_{2}^{\dag}\hat{a}_{2}+S_{z}^{a}\rangle &
\langle\hat{a}_{1}^{\dag}\hat{a}_{1}-\hat{a}_{2}^{\dag}\hat{a}_{2}+iS_{y}^{a}\rangle \\
\langle\hat{a}_{1}^{\dag}\hat{a}_{1}-\hat{a}_{2}^{\dag}\hat{a}_{2}-iS_{y}^{a}\rangle &
\langle\hat{a}_{1}^{\dag}\hat{a}_{1}+\hat{a}_{2}^{\dag}\hat{a}_{2}+S_{z}^{a}\rangle
\end{array}\right),
\label{eq:expandmat}
\end{equation}
which will not in general be diagonal. Leaving aside the fact that a single-party reduced density matrix is only strictly defined in this manner for eigenstates of the number operator and we have seen that, even when we begin our system in number eigenstates, it does not remain in them, we will follow a similar approach and calculate what we will call a ``pseudo entropy" for reasons which will become obvious below.

Rather than using the density matrix defined above, we define an approximate density matrix for the left hand side of the system as
\begin{eqnarray}
R_{a} = \frac{1}{\langle \hat{N}_{a}\rangle}\left(\begin{array}{cc}
\langle \hat{a}_{1}^{\dag}\hat{a}_{1}\rangle & \langle \hat{a}_{1}^{\dag}\hat{a}_{2}\rangle\\
\langle \hat{a}_{2}^{\dag}\hat{a}_{1}\rangle & \langle \hat{a}_{2}^{\dag}\hat{a}_{2}\rangle
\end{array}\right),
\label{eq:nossomat}
\end{eqnarray}
where $\hat{N}_{a}=\hat{a}_{1}^{\dag}\hat{a}_{1}+\hat{a}_{2}^{\dag}\hat{a}_{2}$. This allows us to work with the numbers at each of the two sites, rather than the numbers in modes which combine both sites. It is then an easy matter to calculate a single-particle subsystem pseudo entropy from this matrix,
\begin{equation}
\xi = -\mathrm{Tr}\left(R_{a}\ln R_{a}\right),
\label{eq:entropy}
\end{equation}
which will have a maximum value of $\ln 2 \approx 0.6931$ when the atoms are equally distributed throughout the wells, which is statistically the most probable situation. In Fig.~\ref{fig:energyandentropy} we see that our heuristic entropy does approach this value, although not monotonically. This suggests that, while it is formally wrong to think of Eq.~(\ref{eq:nossomat}) as being a density matrix, it is only approximately wrong when it comes to calculating the single-particle subsystem entropies and could prove to be a useful experimental measure. Indeed, the quantities used to construct this matrix can all be measured experimentally using the techniques developed by Ferris \etal~\cite{Andyhomo}. What we have also seen here that was not visible in the analysis used by Strzys and Anglin is that the details of the approach to the final dynamical equilibrium state depend strongly on the initial quantum states. 

\section{Relaxation to equilibrium}
\label{sec:relax}

The relaxation to equilibrium of closed quantum systems is an important topic of study, as seen in, for example~\cite{thermal1,thermal2,thermal3}, with a beautiful experiment by Kinoshita \etal~\cite{Kinoshita}  having shown that relaxation to equilibrium does not happen in a trapped one dimensional Bose gas. This was not unexpected for a one dimensional untrapped Bose gas with point interactions, which is known to be an integrable system, but it had been thought that practical features such as the harmonic trap and imperfectly point-like interactions would compromise the integrability and the system would relax. Our system is not integrable and is therefore able to equilibriate at zero temperature, without any interactions with a thermal cloud or other reservoir.  However, to the best of our knowledge, there is as yet no consensus on the mechanism by which this happens.

\begin{figure}
\begin{center}
\includegraphics[width=0.6\columnwidth]{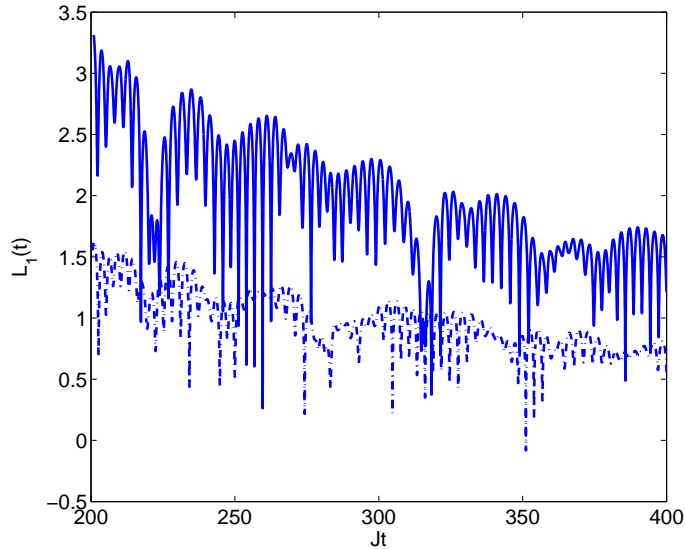}
\end{center}
\caption{(colour online) The Lyupanov exponent $L_{1}(t)$ for the same parameters as used in Fig.~\ref{fig:energyandentropy}. The solid line has one atom added to well $a_{1}$ and one subtracted from $b_{2}$ as the perturbation, while the dash-dotted line has a difference of two atoms.}
\label{fig:Lyapunov}
\end{figure}

For the thermalisation of quantum systems at finite temperature, one proposal is the eigenstate thermalisation hypothesis (ETH), in which every eigenstate of the Hamiltonian implicitly contains a thermal state~\cite{srednicki,thermal3}. Srednicki, when introducing this hypothesis, claimed that a necessary condition was the validity of Berry's conjecture~\cite{Berry}, which is expected to hold for systems which exhibit classical chaos in at least a large majority of the classical phase space. It is a simple matter to calculate effective Lyapunov exponents for the $\alpha_{j}$ in this system, as long as we restrict the inital region of the phase space 
to that close to the initial conditions we have used for well occupations, and thus determine whether classical chaos could be present. A Lyapunov exponent for each well can be defined as
\begin{equation}
L_{j} = \lim_{\tau\rightarrow\infty}\frac{1}{\tau}\frac{\ln\left(\delta\alpha_{j}(\tau)\right)}{\delta\alpha_{j}(0)},
\label{eq:Lyapdef}
\end{equation}
where
\begin{equation}
\delta\alpha_{j}(\tau)=|\alpha_{j}^{(1)}(\tau)-\alpha_{j}^{(2)}(\tau)|,\:\:\:j=1,2,3,4,
\label{eq:deltadef}
\end{equation}
where $\alpha_{j}^{(2)}$ is an initial condition slightly perturbed from $\alpha_{j}^{(1)}$. In practice, we obviously cannot integrate the equations for infinite time, so we integrate the coupled GPE type equations over a reasonably long time and look at the development of $\delta\alpha_{j}(t)$ and hence $L_{j}(t)$. What we found was that the system was stable for initial distributions close to equal numbers in each well, but that, for the initial numbers used in Fig.~\ref{fig:energyandentropy}, it was unstable and therefore chaotic, meaning that it satisfies the criterion for Berry's conjecture to be applicable. The Lyapunov exponents as a function of time are shown in Fig.~\ref{fig:Lyapunov} for this unstable configuration.

The fact that the classical equivalent of our system is chaotic for some initial conditions and stable for others suggests that it may provide a useful laboratory for the investigation of thermalisation in closed quantum systems. It also explains why we do not expect to see full collapses and revivals in the quantum system for arbitrary initial conditions, in contrast to those found for twin wells, where there are as many constants of the motion as there are equations of motion. Although it is prohibitively difficult to calculate either the eigenstates of the system or the full density matrix, we have developed practical alternatives which should be experimentally measurable. Further investigation of this model in terms of zero temperature thermalisation will be a subject of future study.

\section{Conclusions}

We have used stochastic integration in the truncated Wigner representation to examine the four-mode Bose-Hubbard model proposed by Strzys and Anglin as a model for heat transfer, in parameter regimes not considered by the original authors. The inclusion of more quantum effects in our analysis shows that this system is not well described by linearisation of fluctuations about classical solutions, and moreover, that the initial quantum states used have a qualitative effect on the subsequent dynamics. The semi-classical analysis predicts that the system will exhibit First Law phenomenology, with continuing oscillations between different types of energy, and will not exhibit irreversible spontaneous processes which would result in an increase of the system entropy. Our quantum analysis suggests that the oscillations will not necessarily be persistent and that an analogue of irreversible processes takes place which does lead to an increase in entropy, with the system therefore exhibiting both diffusive and oscillatory behaviours.

We suggest that the model may not be a good one for heat transfer, for the reason that it is not at all obvious what the equivalent of temperature is for this system. As temperature will flow from a hotter to colder object, it cannot be the number of atoms in a well, as the overall flow of atoms is at times from a less occupied to a more occupied well. However, the freedom of parameter choices in the initial conditions and the presence of both stable and unstable regions of the initial classical phase space suggest that it will be a highly tractable model for closed system quantum thermalisation.

\section*{Acknowledgments}
This research was supported by the Australian Research Council under the Centres of Excellence Program.

\end{document}